\documentclass[aps,prd,reprint,superscriptaddress,amsmath,amsfonts,amssymb]{revtex4-2}
\usepackage[T1]{fontenc}
\DeclareUnicodeCharacter{2212}{-}
\usepackage[spanish,USenglish]{babel}
\usepackage[colorlinks,citecolor=blue,urlcolor=blue,linkcolor=blue]{hyperref}
\usepackage{braket}
\usepackage{mathtools}
\usepackage{enumerate}
\usepackage{bm}
\usepackage{pgf}
\usepackage{mathrsfs}
\usepackage[caption=false]{subfig}

\newcommand{\schr}{Schr\"odinger}

\DeclareMathAlphabet{\mathbbold}{U}{bbold}{m}{n}

\DeclareMathOperator{\diag}{diag}

\begin{document}

\title{Coupled-channel meson-meson scattering in the diabatic framework}
\author{R. Bruschini}
\email{roberto.bruschini@ific.uv.es}
\affiliation{\foreignlanguage{spanish}{Unidad Te\'orica, Instituto de F\'isica Corpuscular (Universidad de Valencia--CSIC), E-46980 Paterna (Valencia)}, Spain}
\author{P. Gonz\'alez}
\email{pedro.gonzalez@uv.es}
\affiliation{\foreignlanguage{spanish}{Unidad Te\'orica, Instituto de F\'isica Corpuscular (Universidad de Valencia--CSIC), E-46980 Paterna (Valencia)}, Spain}
\affiliation{\foreignlanguage{spanish}{Departamento de F\'isica Te\'orica, Universidad de Valencia, E-46100 Burjassot (Valencia)}, Spain}

\begin{abstract}
We apply the diabatic framework, a QCD-based formalism for the unified study of quarkoniumlike systems in terms of heavy quark-antiquark and open-flavor meson-meson components, to the description of coupled-channel meson-meson scattering. For this purpose, we first introduce a numerical scheme to find the solutions of the diabatic \schr{} equation for energies in the continuum, then we derive a general formula for calculating the meson-meson scattering amplitudes from these solutions. We thus obtain a completely nonperturbative procedure for the calculation of open-flavor meson-meson scattering cross sections from the diabatic potential, which is directly connected to lattice QCD calculations. A comprehensive analysis of various elastic cross sections for open-charm and open-bottom meson-meson pairs is performed in a wide range of the center-of-mass energies. The relevant structures are identified, showing a spectrum of quasi-conventional and unconventional quarkoniumlike states. In addition to the customary Breit-Wigner peaks we obtain nontrivial structures such as threshold cusps and minimums. Finally, our results are compared with existing data and with results from our previous bound-state--based analysis, finding full compatibility with both.
\end{abstract}

\maketitle

\section{Introduction}

Ever since the discovery of the $\chi_{c1}(3872)$ back in 2003 \cite{Cho03}, many quarkoniumlike meson states whose properties defy the conventional description as heavy quark-antiquark bound states have been discovered \cite{PDG20}. The mass of these unconventional states generally lies close below or above the lowest open-flavor threshold with the same quantum numbers. This suggests a possible relevant role of open-flavor meson-meson components. As a matter of fact many theoretical descriptions incorporating meson-meson degrees of freedom have been attempted. These include (but are not limited to) effective field theories, molecular models, and ``unquenched'' quark models. We refer the reader to \cite{Bra11,Bod13,Chen16,Hosaka16,Dong17,Esp17,Leb17,Guo18,Olsen18,Liu19,Yuan19, Bra20} and references therein for a more comprehensive review on the present theoretical and experimental landscape.

In recent years, new theoretical input from lattice QCD has sparked further progress. In particular, the static energy levels for a bottomonium $b \bar{b}$ configuration mixing with one or two open-bottom meson-meson configurations have been calculated in lattice QCD \cite{Bal05,Bul19}, and significant progress  has been made in the understanding of the properties of some charmoniumlike states near open-charm meson-meson thresholds \cite{Bal11,Bal18}. These results have been used to study the effect of up to two spin-averaged thresholds on the bottomonium spectrum \cite{Bic20,Bic21}. A more general framework for the interaction of a $Q \bar{Q}$ pair (where $Q$ stands for a heavy quark, $b$ or $c$) with an arbitrary number of meson-meson channels has been obtained through the implementation of the diabatic approach in QCD \cite{Bru20,Bru21,DiaBot}. This type of approach, first developed in molecular physics \cite{Bae06}, has been applied to calculate the mass spectrum and OZI-allowed strong decay widths of charmoniumlike \cite{Bru20,Bru21} and bottomoniumlike \cite{DiaBot} states.

In the diabatic framework the dynamics, including the $Q$-$\bar{Q}$ interaction and the $Q\bar{Q}$--meson-meson mixing induced by string breaking, is completely described by a potential matrix whose elements are directly related to the static energy levels calculated in lattice QCD. This diabatic potential matrix is then plugged into a multichannel \schr{} equation involving all the $Q\bar{Q}$ and meson-meson components for a given set of $J^{PC}$ quantum numbers. Then, the solutions of this diabatic \schr{} equation allow for the description of quarkoniumlike meson states within the same $J^{PC}$ family \cite{Bru20}. Specifically, bound state solutions for energies below the lowest open-flavor $J^{PC}$ threshold can be directly assigned to quarkoniumlike meson states. In contrast, solutions for energies above one or more thresholds, containing as many free-wave meson-meson components, cannot be directly assigned to physical mesons. Indeed, a dedicated formalism has to be developed in order to obtain a proper physical description.

As an alternative, one may try to avoid this difficulty following a bound state approximation even for energies above threshold. This is, quarkoniumlike meson states in this energy region can be assigned to bound state solutions of a reduced set of \schr{} equations where the coupling with open thresholds is neglected. Then, decay widths and mass shifts induced from the coupling of these bound states with the continuum of free meson-meson states can be evaluated using a procedure well known in nuclear physics and hadron spectroscopy \cite{Fan61, Eic80}. Actually, we have previously followed this approximation for a spectral analysis of charmoniumlike and bottomoniumlike mesons \cite{Bru20,Bru21,DiaBot}. Notwithstanding the good results obtained, this approximation presents some shortcomings. On the one hand, there is an ambiguity regarding the number of bound states for a given $J^{PC}$ related to the possibility of choosing different sets of neglected thresholds when calculating some of them. In \cite{Bru20,Bru21,DiaBot} this ambiguity was obviated \textit{ad hoc} by assuming a one-to-one correspondence between calculated bound states and Cornell states. On the other hand, mixing between different $Q\bar{Q}$ partial waves induced through their coupling to the same meson-meson thresholds is not properly taken into account in the case of open thresholds. Finally, there may be physical mesons escaping the bound state approximation.

In order to overcome these shortcomings we develop in this article a formalism to describe quarkoniumlike mesons from the continuum solutions of the diabatic \schr{} equation above threshold. More precisely, we solve numerically the diabatic \schr{} equation for energies above threshold. Then we decompose the asymptotic limit of each solution, which consists of one or more free waves in various meson-meson channels, as a superposition of stationary meson-meson scattering states. From this decomposition we obtain the on-shell $\mathrm{S}$-matrix for the coupled-channel meson-meson scattering process. Finally, we identify quarkoniumlike meson states as resonances in the calculated scattering cross sections as a function of the energy. From this analysis we recover the quarkoniumlike meson states obtained in the bound state approximation as well as identify additional structures in the cross sections.

It should be pointed out that numerous treatments of the effects of coupled channels in the quarkoniumlike meson spectrum can be found in the literature, see for instance \cite{Eic80, Eic04, Eic06, vanB83, vanB21, Ono84, Bar05, Bar08, Pen07, Dan10, Dan12, Fer13, Fer14, Zhou14, Ort18}. A major difference of these treatments with respect to our diabatic approach is the use of a phenomenological mixing potential (mostly through a $^{3}\!P_0$ model) instead of an interaction based on lattice QCD. Furthermore in some of these treatments a perturbative approach, whose validity is questionable (see Appendix in \cite{Bru21}), has been employed. Such differences make rather difficult a direct comparison between these studies and ours. Somewhat closer to our approach is a quite recent first attempt to study $J^{PC}=0^{++},2^{++}$ charmoniumlike resonances in coupled $D\bar{D}$-$D_{s}\bar{D}_{s}$ scattering on the lattice \cite{Prel21}. However, the several simplifying  assumptions needed for this first lattice investigation of the coupled-channel system  also prevent a direct comparison with our results.

The contents of this article are organized as follows. In Sec.~\ref{hamsec} we briefly revisit the diabatic approach in QCD and write down the diabatic \schr{} equation. Section~\ref{numsec} treats the numerical method used to solve the diabatic \schr{} equation for continuum energies above the lowest open-flavor threshold. The theoretical framework used to describe coupled-channel meson-meson scattering is detailed in Sec.~\ref{scatsec}, where we derive the formula for calculating the $\mathrm{S}$-matrix. The calculated elastic cross sections are briefly discussed in Sec.~\ref{ressec}. Then in Sec.~\ref{endsec} we summarize our main findings.

\section{Diabatic \schr{} equation\label{hamsec}}

The diabatic approach in QCD is a recently proposed theoretical framework aiming at a unified description of both conventional and unconventional quarkoniumlike states based on lattice QCD results. Its construction, as well as its differences from other commonly used approaches in heavy-meson spectroscopy, has been explained extensively in \cite{Bru20}. In this section we briefly recap its main features.

The dynamics is governed by the diabatic \schr{} equation
\begin{equation}
(K + V) \ket{\Psi} = E \ket{\Psi}
\label{deq}
\end{equation}
where $\ket{\Psi}$ is the state of the system containing one $Q\bar{Q}$ and $N$ meson-meson components $M_{1}^{(i)} \bar{M}_{2}^{(i)}$ with $i = 1, \dots, N$, $E$ the c.m.\ energy, $K$ the kinetic energy operator, and $V$ the diabatic potential operator. If we conveniently represent $\ket{\Psi}$ as a column vector
\begin{equation}
\ket{\Psi} =
\begin{pmatrix}
\ket{\psi^{(0)}}	\\
\ket{\psi^{(1)}}	\\
\vdots 		\\
\ket{\psi^{(N)}}	\\
\end{pmatrix}
\end{equation}
where $\ket{\psi^{(0)}}$ and $\ket{\psi^{(i)}}$ are the states of the $Q\bar{Q}$ and $M_{1}^{(i)}\bar{M}_{2}^{(i)}$ components respectively, then we can represent the kinetic energy operator as a matrix
\begin{equation}
K =
\begin{pmatrix}
\frac{\mathrm{p}^{2}}{2\mu^{(0)}}		&										&		&							\\
									& \frac{\mathrm{p}^{2}}{2\mu^{(1)}}			& 		&							\\
									&										& \ddots 	&							\\
									&										&		& \frac{\mathrm{p}^{2}}{2\mu^{(N)}}	\\
\end{pmatrix}
\label{kinetic}
\end{equation}
with $\mu^{(0)}$ and $\mu^{(i)}$ respectively the reduced mass of the $Q\bar{Q}$ and $M_{1}^{(i)} \bar{M}_{2}^{(i)}$ component and $\mathrm{p}^{2}$ the squared relative momentum operator. Accordingly, in this notation the diabatic potential operator reads
\begin{equation}
V =
\begin{pmatrix}
V^{(0)}					& V_{\textup{mix}}^{(1)}			& \hdots	& V_{\textup{mix}}^{(N)}			\\
V_{\textup{mix}}^{(1) \dagger} 	& V^{(1)}						& 		&							\\
\vdots 					&							& \ddots 	&							\\
V_{\textup{mix}}^{(N) \dagger}	&							&		& V^{(N)}						\\
\end{pmatrix}
\label{dpot}
\end{equation}
where vanishing matrix elements have been omitted.

By projecting on the relative position space we have
\begin{equation}
\int \mathrm{d} \bm{r}^{\prime} \braket{\bm{r} \rvert K + V \lvert \bm{r}^{\prime}} \braket{\bm{r}^{\prime} \vert \Psi} = E \braket{\bm{r} \vert \Psi}
\label{posdeq}
\end{equation}
where the wave function $\braket{\bm{r} \vert \Psi}$ is written as
\begin{equation}
\braket{\bm{r} \vert \Psi} =
\begin{pmatrix}
\psi^{(0)}(\bm{r})	\\
\psi^{(1)}(\bm{r})	\\
\vdots 			\\
\psi^{(N)}(\bm{r})	\\
\end{pmatrix}
\label{wfunc}
\end{equation}
with $\psi^{(0)}(\bm{r}) \equiv \braket{\bm{r} \vert \psi^{(0)}}$ and $\psi^{(i)}(\bm{r}) \equiv \braket{\bm{r} \vert \psi^{(i)}}$ respectively the wave function of the $Q\bar{Q}$ and $M_{1}^{(i)} \bar{M}_{2}^{(i)}$ component.

It is important to realize that, since we do not project explicitly on spin, each component of the wave function \eqref{wfunc} is intended as a spin vector. In practice, since the diabatic potential must incorporate the symmetries of QCD, we impose $J^{PC}$ conservation. Therefore each solution to the diabatic \schr{} equation is labeled by its $J^{PC}$ and $m_{J}$ quantum numbers, with $J$ and $m_{J}$ the total angular momentum and its projection, $P$ the parity, and $C$ the charge-conjugation parity. Then the wave function components are more conveniently expressed in terms of the irreducible tensors of order $J$:
\begin{equation}
\mathcal{Y}_{l, s}^{J, m_{J}}(\bm{\hat{r}}) \equiv \braket{\bm{\hat{r}} \vert l, s, J, m_{J}} = \sum_{m_{l}, m_{s}} C_{l, s, J}^{m_{l}, m_{s}, m_{J}} Y_{l}^{m_{l}}(\bm{\hat{r}}) \xi_{s}^{m_{s}}
\end{equation}
satisfying
\begin{equation}
\int \mathrm{d} \bm{\hat{r}} \mathcal{Y}_{l, s}^{J, m_{J} \dagger}(\bm{\hat{r}}) \mathcal{Y}_{l^{\prime}, s^{\prime}}^{J^{\prime}, m_{J^{\prime}}}(\bm{\hat{r}}) = \delta_{J J^{\prime}} \delta_{m_{J} m_{J^{\prime}}} \delta_{l l^{\prime}} \delta_{s s^{\prime}}
\end{equation}
where $\bm{r} = r \bm{\hat{r}}$, $C_{l, s, J}^{m_{l}, m_{s}, m_{J}}$ are Clebsch-Gordan coefficients, $Y_{l}^{m_{l}}(\bm{\hat{r}})$ spherical harmonics, and $\xi_{s}^{m_{s}}$ spin vectors. Concretely, the wave function components are expanded as
\begin{align}
\psi_{J^{PC}, m_{J}}^{(0)}(\bm{r}) &= \sum_{t} R_{J^{PC}; t}^{(0)}(r) \mathcal{Y}_{l_{t}^{(0)},s_{t}^{(0)}}^{J,m_{J}}(\bm{\hat{r}}) \label{qqwfuncexp} \\
\psi_{J^{PC}, m_{J}}^{(i)}(\bm{r}) &=\sum_{k} R_{J^{PC}; k}^{(i)}(r) \mathcal{Y}_{l_{k}^{(i)},s_{k}^{(i)}}^{J,m_{J}} (\bm{\hat{r}}) \label{mesmeswfuncexp}
\end{align}
where $t$ and $k$ label the distinct $(l^{(0)}, s^{(0)})$ and $(l^{(i)}, s^{(i)})$ partial waves coupling to $J^{PC}$ in the $Q\bar{Q}$ and $M_{1}^{(i)} \bar{M}_{2}^{(i)}$ configuration respectively, while $R_{J^{PC}; t}^{(0)}(r)$ and $R_{J^{PC}; k}^{(i)}(r)$ stand for radial wave functions.

The kernel (i.e, the position space representation) of the kinetic energy operator reads
\begin{equation}
\braket{\bm{r} \rvert K \lvert \bm{r}^{\prime}} = \delta(\bm{r}^{\prime} - \bm{r})
\begin{pmatrix}
-\frac{\nabla^{2}}{2\mu^{(0)}}		&										&		&							\\
									& -\frac{\nabla^{2}}{2\mu^{(1)}}			& 		&							\\
									&										& \ddots 	&							\\
									&										&		& -\frac{\nabla^{2}}{2\mu^{(N)}}	\\
\end{pmatrix}.
\label{kmat}
\end{equation}

As for the potential energy the diagonal term $V^{(0)}$, describing the $Q$-$\bar{Q}$ interaction, is directly connected to quenched lattice QCD results for the static light field energy \cite{Bal01}. It is represented by the kernel
\begin{equation}
\braket{\bm{r} \rvert V^{(0)} \lvert \bm{r}^{\prime}} = \delta(\bm{r}^{\prime} - \bm{r}) V_{\textup{C}}(r)
\label{v0exp}
\end{equation}
where $V_{\textup{C}}(r)$ corresponds to the Cornell potential
\begin{equation}
V_{\textup{C}}(r) = \sigma r - \frac{\chi}{r} + m_{Q} + m_{\bar{Q}} - \beta
\label{cpot}
\end{equation}
with $\sigma$, $\chi$, $m_{Q}$ ($m_{\bar{Q}}$), and $\beta$ respectively the string tension, color Coulomb strength, heavy quark (antiquark) mass, and a constant. The other diagonal terms read
\begin{equation}
\braket{\bm{r} \rvert V^{(i)} \lvert \bm{r}^{\prime}} = \delta(\bm{r}^{\prime} - \bm{r}) T^{(i)}
\label{viexp}
\end{equation}
with $T^{(i)}$ the threshold value
\begin{equation}
T^{(i)} = m_{M_{1}^{(i)}} + m_{\bar{M}_{2}^{(i)}}
\label{mmpot}
\end{equation}
where $m_{M_{1}^{(i)}}$ and $m_{\bar{M}_{2}^{(i)}}$ are the masses of the corresponding mesons.

As for the nonvanishing offdiagonal elements $V_{\textup{mix}}^{(i)}$, governing the $Q\bar{Q}$-$M_{1}^{(i)} \bar{M}_{2}^{(i)}$ mixing due to string breaking, they can be determined \textit{ab initio} from the static energy levels calculated in lattice QCD. This is done by expanding their kernel in $J^{PC}$-conserving tensors,
\begin{multline}
\braket{\bm{r} \rvert V_{\textup{mix}}^{(i)} \lvert \bm{r}^{\prime}} = \delta(r^{\prime} - r) \\
V_\textup{mix}^{(i)}(r) \sum_{J^{PC}, m_{J}} \sum_{t^{\prime}, k^{\prime}} \mathcal{Y}_{l_{t^{\prime}}^{(0)}, s_{t^{\prime}}^{(0)}}^{J,m_{J}}(\bm{\hat{r}}) \mathcal{Y}_{l_{k^{\prime}}^{(i)}, s_{k^{\prime}}^{(i)}}^{J,m_{J} \dagger} (\bm{\hat{r}}^{\prime})
\label{mixpotexp}
\end{multline}
where for simplicity we have assumed the radial mixing potentials $V_\textup{mix}^{(i)}(r)$ to be independent of the $l$, $s$, and $J^{PC}$ quantum numbers. Thus, using Eqs.~\eqref{kmat}-\eqref{mixpotexp}, the diabatic \schr{} equation \eqref{posdeq} reduces to a multichannel radial \schr{} equation with a radial diabatic potential matrix written in terms of $V_\textup{C}(r)$, $T^{(i)}$, and $V_\textup{mix}^{(i)}(r)$. It is then possible to determine $V_\textup{mix}^{(i)}(r)$ by virtue of the direct correspondence between the static energy levels and the eigenvalues of the diabatic potential matrix \cite{Bae06}. Our educated guess \cite{Bru20},
\begin{equation}
\bigl\lvert{V_{\textup{mix}}^{(i)}(r)}\bigr\rvert = \frac{\Delta^{(i)}}{2} \exp \biggl\{- \frac{1}{2} \biggl( \frac{V_{\textup{C}}(r)-T^{(i)}}{\sigma \rho^{(i)}} \biggr)^{2}\biggr\}
\label{mixpot}
\end{equation}
where the factors $\Delta^{(i)}$ and $\rho^{(i)}$ respectively represent the strength and mixing length scale of the mixing, is based on the static energy levels recently calculated in unquenched lattice QCD \cite{Bal05,Bul19}. The physical reasons behind this parametrization have been detailed in \cite{Bru20}, which we refer to.

It may be instructive to derive explicitly the multichannel radial equation in the simplest possible example of a $Q\bar{Q}$ component interacting with a single meson-meson component $M_{1}^{(1)} \bar{M}_{2}^{(1)}$, each with only one partial wave contributing to $J^{PC}$. In this case Eq.~\eqref{posdeq} reads
\begin{widetext}
\begin{equation}
\begin{pmatrix}
(- \frac{\nabla^{2}}{2 \mu^{(0)}} + V_\textup{C}(r)) \psi^{(0)}_{J^{PC}, m_{J}}(\bm{r}) \\
(- \frac{\nabla^{2}}{2 \mu^{(1)}} + T^{(1)}) \psi^{(1)}_{J^{PC}, m_{J}}(\bm{r})
\end{pmatrix}
+
\begin{pmatrix}
V_\textup{mix}^{(1)}(r) \mathcal{Y}_{l_{1}^{(0)}, s_{1}^{(0)}}^{J,m_{J}}(\bm{\hat{r}}) \int \mathrm{d} \bm{\hat{r}}^{\prime} \mathcal{Y}_{l_{1}^{(1)}, s_{1}^{(1)}}^{J,m_{J} \dagger} (\bm{\hat{r}}^{\prime}) \psi^{(1)}_{J^{PC}, m_{J}}(r \bm{\hat{r}}^{\prime}) \\
V_\textup{mix}^{(1)}(r) \mathcal{Y}_{l_{1}^{(1)}, s_{1}^{(1)}}^{J,m_{J}}(\bm{\hat{r}}) \int \mathrm{d} \bm{\hat{r}}^{\prime} \mathcal{Y}_{l_{1}^{(0)}, s_{1}^{(0)}}^{J,m_{J} \dagger} (\bm{\hat{r}}^{\prime}) \psi^{(0)}_{J^{PC}, m_{J}}(r \bm{\hat{r}}^{\prime})
\end{pmatrix}
= E
\begin{pmatrix}
\psi^{(0)}_{J^{PC}, m_{J}}(\bm{r}) \\
\psi^{(1)}_{J^{PC}, m_{J}}(\bm{r})
\end{pmatrix}.
\end{equation}
If we now substitute Eqs.~\eqref{qqwfuncexp} and \eqref{mesmeswfuncexp}, multiply by
$
\begin{pmatrix}
\mathcal{Y}_{l_{1}^{(0)}, s_{1}^{(0)}}^{J,m_{J}}(\bm{\hat{r}})	& \mathcal{Y}_{l_{1}^{(1)}, s_{1}^{(1)}}^{J,m_{J}}(\bm{\hat{r}})
\end{pmatrix}$,
and integrate over the solid angle $\bm{\hat{r}}$, we get the multichannel radial equation
\begin{equation}
\left[
\begin{pmatrix}
- \frac{1}{2 \mu^{(0)}} \frac{\mathrm{d}^{2}}{\mathrm{d} r^{2}} + \frac{l_{1}^{(0)} (l_{1}^{(0)} + 1)}{2 \mu^{(0)} r^{2}}	& \\
& - \frac{1}{2 \mu^{(1)}} \frac{\mathrm{d}^{2}}{\mathrm{d} r^{2}} + \frac{l_{1}^{(1)} (l_{1}^{(1)} + 1)}{2 \mu^{(1)} r^{2}}
\end{pmatrix}
+
\begin{pmatrix}
V_\textup{C}(r)		& V_\textup{mix}^{(1)}(r)	\\
V_\textup{mix}^{(1)}(r)	& T^{(1)}
\end{pmatrix}\right]
\begin{pmatrix}
u_{J^{PC}; 1}^{(0)}(r) \\
u_{J^{PC}; 1}^{(1)}(r)
\end{pmatrix}
=
E \begin{pmatrix}
u_{J^{PC}; 1}^{(0)}(r) \\
u_{J^{PC}; 1}^{(1)}(r)
\end{pmatrix}
\label{radeq}
\end{equation}
\end{widetext}
where we have introduced the reduced radial wave functions $u_{J^{PC}; 1}^{(0)}(r) \equiv r R_{J^{PC}; 1}^{(0)}(r)$ and $u_{J^{PC}; 1}^{(1)}(r) \equiv r R_{J^{PC}; 1}^{(1)}(r)$.

The generalization to any number of partial waves and an arbitrary number of meson-meson components is straightforward by considering each $u_{J^{PC}; t}^{(0)}(r)$ and each $u_{J^{PC}; k}^{(i)}(r)$ as an independent component of the eigenfunction.

\begin{figure}
\subfloat[][Light meson exchange.\label{mesexch}]{\includegraphics[width=.45\columnwidth]{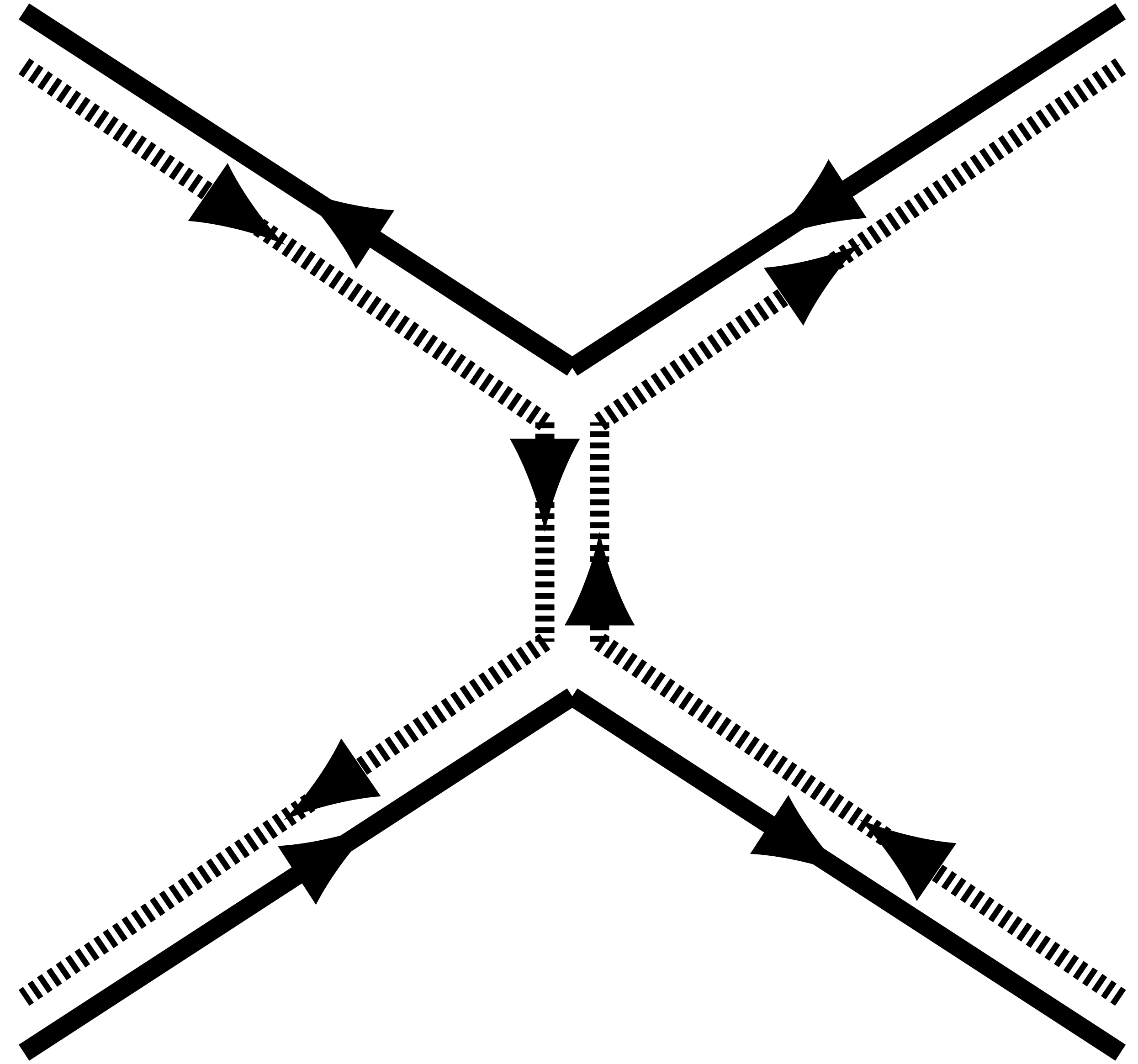}} \quad
\subfloat[][Mixing with $Q\bar{Q}$.\label{qqexch}]{\includegraphics[width=.45\columnwidth]{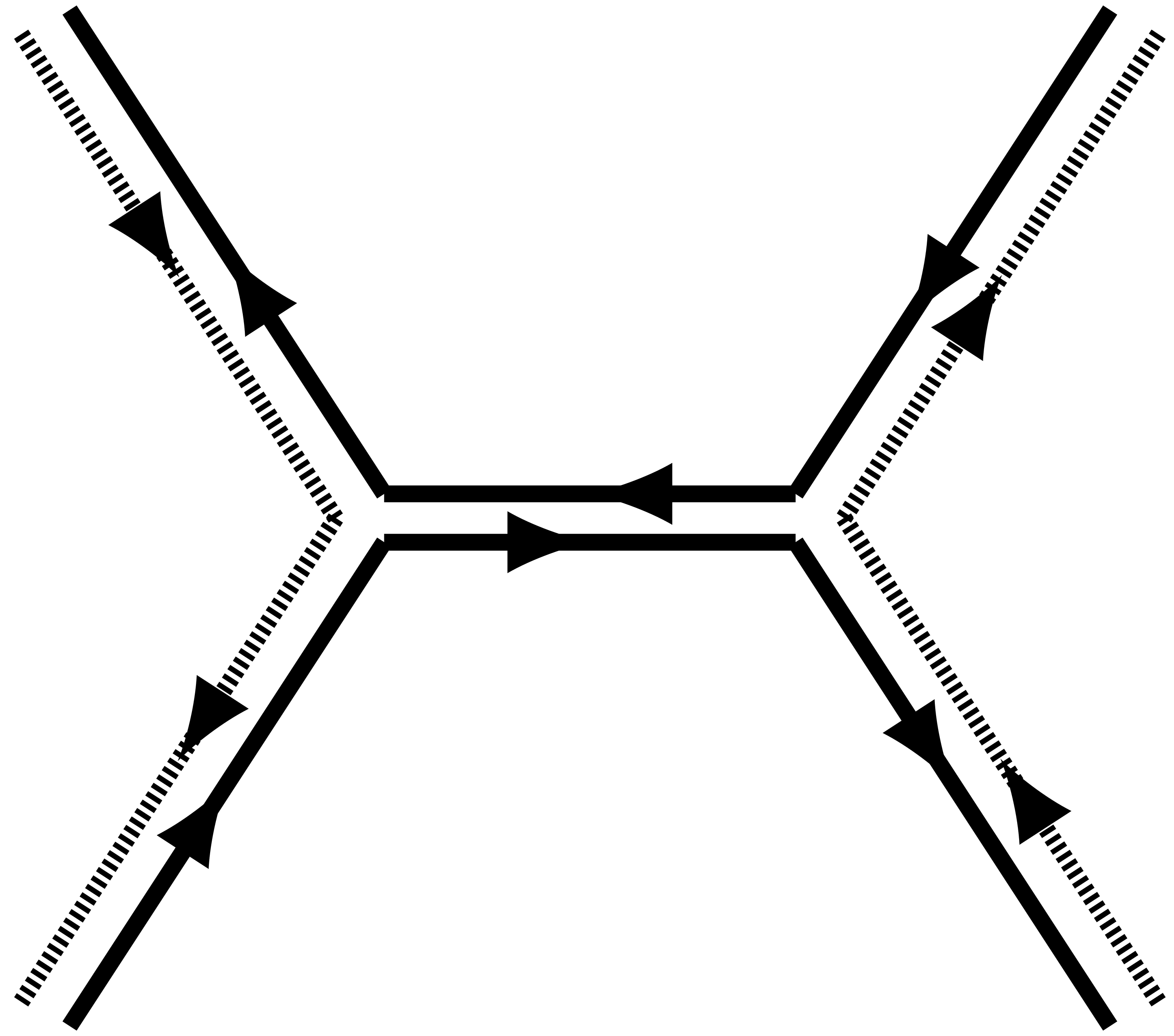}}
\caption{Diagrammatic representation of possible meson-meson interactions. Solid and dashed lines correspond to heavy and light quarks, respectively.}
\end{figure}

It has to be remarked that by using Eqs.~\eqref{dpot}, \eqref{viexp}, and \eqref{mmpot} we are neglecting all meson-meson interactions from light meson exchange depicted in Fig.~\ref{mesexch}. Instead, the diabatic potential incorporates a meson-meson interaction mediated by $Q\bar{Q}$ depicted in Fig.~\ref{qqexch}. Then our approximation could be regarded as the assumption that other interactions (like light meson exchange potentials) are subdominant with respect to the diabatic $Q\bar{Q}$-mediated ones. Although similar approximations for the potential matrix are commonly used in this context (compare for example with Eq.~(7) of Ref.~\cite{Bul19}), this assumption may be debatable. In fact there have been effective field theory studies, see for instance \cite{Alb13, Hid13}, where the dynamical energy levels have been described through a dominant contact meson-meson interaction. Therefore our treatment should be considered as an exploratory research on the effect of the mixing term. One should keep in mind though that additional contributions to the potential might be implicitly taken into account through the effective values of the parameters in the mixing potential.

For energies below the lowest $J^{PC}$ threshold the spectrum of solutions to the diabatic \schr{} equation consists in a certain number of discrete energies with wave functions representing properly normalizable states. Then, a discrete solution with energy $E$ can be assigned to a quarkoniumlike meson with mass $E$. Concretely, solutions for energies far below the lowest threshold, containing a very dominant $Q\bar{Q}$ component, differ very little from the quark model solutions obtained from the Cornell potential. They correspond to the lowest-lying conventional quarkonium (charmonium and bottomonium) states observed experimentally. In contrast, solutions with a relevant meson-meson component may appear close below the lowest threshold, as may be the case for the charmoniumlike meson $\chi_{c1}(3872)$ \cite{Bru20}.

For energies above one or more thresholds the solutions, containing one or more free-wave meson-meson components, cannot be directly assigned to physical mesons. Instead, they have a more natural interpretation in terms of stationary meson-meson scattering states. More precisely, the diabatic \schr{} equation for any energy $E$ above threshold can be used to describe the $J^{PC}$ contribution to a meson-meson scattering process at c.m.\ energy $E$, where the meson-meson interaction is mediated by $Q\bar{Q}$ (see Fig.~\ref{qqexch}). From this interpretation, quarkoniumlike mesons observed experimentally can be properly assigned to meson-meson scattering resonances.

\section{Numerical solution above threshold\label{numsec}}

In this section we briefly revisit the finite difference method for the solution of the diabatic \schr{} equation and show how the imposition of appropriate boundary conditions can be used to obtain a set of independent solutions for each energy above threshold.

In order to keep the discussion both simple and general, we start by examining the simplest case of a single radial \schr{} equation
\begin{equation}
-\frac{1}{2 \mu} u^{\prime\prime}(r) + \biggl( \frac{l (l + 1)}{2 \mu r^{2}} + V(r) - E \biggr) u(r) = 0
\label{singleq}
\end{equation}
where $V(r)$ is some spherical potential, $u(r)$ a reduced radial wave function, $\mu$ a reduced mass, and $l$ a relative orbital angular momentum.

The finite difference method consists in discretizing the radial configuration space in a lattice of equally spaced points $r_{\mathrm{n}}$. This gives rise to a matrix \schr{} equation whose solution yields the numerical wave function $u_{\mathrm{n}}\equiv u(r_{\mathrm{n}})$.

Discretization of the centrifugal and potential energy terms in \eqref{singleq} is straightforward, while that of the second derivative proceeds in two steps. The first step is to introduce the $\mathcal{O}(d^{2})$ approximation for the second derivative
\begin{equation}
u_{\mathrm{n}}^{\prime\prime} \approx \frac{u_{\mathrm{n} + 1} - 2 u_{\mathrm{n}} + u_{\mathrm{n} - 1}}{d^{2}}
\label{secder}
\end{equation}
where $d$ is the discretization step. Notice however that this expression cannot be applied at the origin and extreme of the lattice, given respectively by $r_0 = 0$ and $r_\textup{extr} \equiv r_{\mathrm{N} + 1} = (\mathrm{N} + 1) d$ where $\mathrm{N}$ is the number of points in its interior. Thus the second step is to impose for these points Dirichlet boundary conditions $u_{0} = b_{0}$ and $u_{\mathrm{N} + 1} = b_\textup{extr}$, so the second derivative for $\mathrm{n} = 1$ and $\mathrm{n} = \mathrm{N}$ reads
\begin{align}
u^{\prime\prime}_{1} &\approx \frac{u_{2} - 2 u_{1}}{d^{2}} + \frac{b_{0}}{d^{2}}, \label{secderzero}\\
u^{\prime\prime}_{\mathrm{N}} &\approx \frac{u_{\mathrm{N} - 1} - 2 u_{\mathrm{N}}}{d^{2}} + \frac{b_{\textup{extr}}}{d^{2}}, \label{secderextr}
\end{align}
while for $\mathrm{n}=2,\dots, \mathrm{N}-1$ it is given by Eq.~\eqref{secder}.

The discretized spherical \schr{} equation is then written in matrix form as 
\begin{equation}
(\mathrm{H} - E) \mathrm{u} = \mathrm{B}
\label{schrgen}
\end{equation}
where $\mathrm{u}$ is the numerical reduced wave function (without the boundary values)
\begin{equation}
\mathrm{u} =
\begin{pmatrix}
u_{1} \\
\vdots \\
u_{\mathrm{N}}
\end{pmatrix},
\end{equation}
$\mathrm{H}$ is the Hamiltonian matrix
\begin{equation}
\mathrm{H} = \mathrm{K} + \mathrm{V}_\textup{eff}
\end{equation}
with $\mathrm{K}$ the tridiagonal kinetic energy matrix
\begin{equation}
\mathrm{K} = - \frac{1}{2 \mu d^{2}}
\begin{pmatrix}
-2	& 1		&	 	&		&		\\
1	& -2		& 1 		&		&		\\
	& \ddots	& \ddots 	& \ddots	&		\\
	&		&	1	& -2		& 1 		\\
	&		&		& 1		& -2		\\
\end{pmatrix},
\label{kinmat}
\end{equation}
$\mathrm{V}_\textup{eff} = \diag(V_\textup{eff}(\mathrm{r}_{1}),\dots,V_\textup{eff}(\mathrm{r}_{\mathrm{N}}))$ where $V_\textup{eff}(r) \equiv V(r) + \frac{l (l + 1)}{2 \mu r^{2}}$, and $\mathrm{B}$ a constant numerical vector related to the boundary conditions, whose only nonzero components are in general the first and the last one:
\begin{equation}
\mathrm{B} = \frac{1}{2 \mu d^{2}}
\begin{pmatrix}
b_{0} \\
\\
b_{\textup{extr}}
\end{pmatrix}.
\end{equation}

If the potential $V(r)$ is either regular or diverges at most as $r^{-1}$ for $r\to 0$, as it is always in our case, the physical boundary condition at the origin is $b_{0}=0$. As for the boundary condition at $r_\textup{extr}$, assuming $r_\textup{extr}$ to be very large, there are two distinct cases:
\begin{enumerate}[i)]
\item if $E < \lim_{r\to\infty}V(r)$, the physical boundary condition is $b_\textup{extr}=0$\label{firstcase};
\item if otherwise $E > \lim_{r\to\infty}V(r)$, any finite $b_\textup{extr}$ is physical\label{secondcase}.
\end{enumerate}
These two scenarios lead to very different treatments of the numerical \schr{} equation.

In the first case we have $\mathrm{B}=0$ and Eq.~\eqref{schrgen} reduces to a secular equation for the Hamiltonian matrix $\mathrm{H}$, whose solution provides the bound state spectrum and the corresponding wave functions.

In the second case the energy spectrum is known \textit{a priori}, being the continuum $E \ge \lim_{r\to\infty}V(r)$, but there is no physical criterion to fix the boundary condition at the extreme. Then for each continuum energy $E$ the wave functions are calculated solving the nonhomogeneous Eq.~\eqref{schrgen} with an arbitrary nonzero boundary condition $b_\textup{extr}\ne 0$. Indeed, as we have fixed the boundary condition at $r_{0}$ there is only one linearly independent solution for each value of $E$, therefore different choices of $b_{\textup{extr}}$ will produce the same wave function up to a global multiplicative factor. Note that the assumption $b_{\textup{extr}} \ne 0$ is not restrictive in numerical applications. In fact, since the energies yielding a nontrivial solution that vanishes at $r_\textup{extr}$ constitute a discrete subset of the continuum (i.e., the energy levels of a particle in a spherical box), the chance of picking accidentally the exact numerical value of one such energy is negligible.

The procedure we have detailed for a single-channel equation can be easily extended to the solution of the diabatic \schr{} equation. We begin by observing that if instead of \eqref{singleq} one has a multichannel radial \schr{} equation (like Eq.~\eqref{radeq} for example), then the discretization procedure yields a system of coupled numerical equations that can be rearranged in the form \eqref{schrgen}, with the tridiagonal matrix $\mathrm{H}$ substituted by a more general Hermitian banded matrix.

Regarding the boundary condition at $r_\textup{extr}$, note that each diabatic channel falls into one of the two aforementioned cases \ref{firstcase}) or \ref{secondcase}) depending on the energy. In particular, $Q\bar{Q}$ and closed meson-meson channels fall into category \ref{firstcase}) whereas open meson-meson channels fall into category \ref{secondcase}). So for energies below the lowest threshold, where all meson-meson channels are closed, all reduced wave function components vanish at ``infinity'', which corresponds to bound state solutions. Otherwise above threshold the boundary condition at $r_\textup{extr}$ becomes a numerical vector whose only nonzero block is that corresponding to open meson-meson channels,
\begin{equation}
b_{\textup{extr}} \rightarrow \mathrm{b}_\textup{extr} =
\begin{pmatrix}
\mathrm{b}_{Q\bar{Q}} \\
\mathrm{b}_\textup{open} \\
\mathrm{b}_\textup{closed}
\end{pmatrix}
\end{equation}
with
\begin{equation}
\mathrm{b}_{Q\bar{Q}} = 0, \quad
\mathrm{b}_\textup{open} =
\begin{pmatrix}
b^{(1)} \\
\vdots 				\\
b^{(\tilde{n})}
\end{pmatrix}, \quad
\mathrm{b}_\textup{closed} = 0
\label{multbcond}
\end{equation}
where $\tilde{n}$ is the total number of partial waves coupling to $J^{PC}$ in the open meson-meson channels.

Finally, notice that one can choose up to $\tilde{n}$ linearly independent boundary conditions $\mathrm{b}_\textup{open}$, yielding as many independent solutions with the same energy. Therefore, for an arbitrary continuum energy one is able to build a basis set of solutions to the diabatic \schr{} equation by solving \eqref{schrgen} with boundary conditions \eqref{multbcond} for $\tilde{n}$ linearly independent numerical vectors $\mathrm{b}_\textup{open}$. In practice, we build and solve the nonhomogeneous linear system \eqref{schrgen} in Python using the NumPy and SciPy libraries \cite{numpy,scipy} and taking the canonical basis of $\mathbb{C}^{\tilde{n}}$ as our set of $\tilde{n}$ linearly independent vectors $\mathrm{b}_\textup{open}$.

For the sake of completeness, we report here that in this paper we have used $d = 10^{-3}$~fm and $r_\textup{extr} = 200$~fm.

\section{Multichannel scattering states\label{scatsec}}

\subsection{Asymptotic solutions above threshold}

Let us now examine in detail the asymptotic $r \to \infty$ behavior of a solution to the diabatic \schr{} equation above threshold, beginning with the simplest possible example of a single open meson-meson channel $M_{1}^{(1)} \bar{M}_{2}^{(1)}$ with only one partial wave $(l_{1}^{(1)}, s_{1}^{(1)})$ coupling to $J^{PC}$.

If there were no mixing, $V_{\textup{mix}}(r) = 0$, the spectrum of the diabatic \schr{} equation would decompose in a discrete spectrum of pure quarkonium states ($\psi^{(1)}(r) = 0$) and a continuum $E \ge T$ of pure meson-meson ($\psi^{(0)}(r) = 0$) free states given by
\begin{equation}
\psi^{(1)}_{J^{PC}, m_{J}}(\bm{r}) = \sqrt{\frac{2}{\pi} \mu^{(1)}p^{(1)}} i^{l_{1}^{(1)}} j_{l_{1}^{(1)}} \bigl(p^{(1)} r\bigr) \mathcal{Y}_{l_{1}^{(1)}, s_{1}^{(1)}}^{J, m_{J}} (\bm{\hat{r}})
\end{equation}
with $j_{l}(p r)$ the $l$th spherical Bessel function of the first kind, $\bm{p}^{(1)} = p^{(1)} \bm{\hat{p}}^{(1)}$ the relative meson-meson momentum with modulus $p^{(1)} = \sqrt{2 \mu^{(1)} (E - T)}$, and $\sqrt{\frac{2}{\pi} \mu^{(1)}p^{(1)}} i^{l_{1}^{(1)}}$ a normalization factor introduced to facilitate the connection with the scattering states (see below). Then, the asymptotic behavior of a continuum solution would be written as
\begin{multline}
\psi^{(1)}_{J^{PC}, m_{J}}(\bm{r}) \simeq \frac{1}{r} \sqrt{\frac{2}{\pi} \frac{\mu^{(1)}}{p^{(1)}}} i^{l_{1}^{(1)}} \\
\sin \Bigl(p^{(1)} r - l_{1}^{(1)} \frac{\pi}{2} \Bigr) \mathcal{Y}_{l_{1}^{(1)}, s_{1}^{(1)}}^{J, m_{J}} (\bm{\hat{r}})
\end{multline}
where we have defined the symbol $\simeq$ for the asymptotic equality relation, meaning that the two sides are equal in the limit $r\to\infty$, and used
\begin{equation}
j_{l}(p r) \simeq \frac{1}{p r} \sin\biggl(p r - l \frac{\pi}{2} \biggr).
\end{equation}

If now we introduce the mixing potential \eqref{mixpot}, the continuum solutions cease being pure meson-meson states as they acquire a $Q \bar{Q}$ component as well.  Notice though that as the mixing potential \eqref{mixpot} goes to zero exponentially for $r\to\infty$, the $Q\bar{Q}$ behaves as a bound component vanishing very quickly in the asymptotic limit. Hence, these mixed free-bound states correspond to a meson-meson pair interacting at short distances through the mixing with $Q\bar{Q}$. The asymptotic behavior of these states is known from elastic scattering theory, see for example Eq.~(11.17) in \cite{Tay72}, as
\begin{multline}
\psi^{(1)}_{J^{PC}, m_{J}}(\bm{r}) \simeq \frac{1}{r} \sqrt{\frac{2}{\pi} \frac{\mu^{(1)}}{p^{(1)}}} i^{l_{1}^{(1)}} e^{i \eta_{J^{PC};1}^{(1)}} \\
\sin \Bigl(p^{(1)} r - l_{1}^{(1)} \frac{\pi}{2} + \eta_{J^{PC};1}^{(1)} \Bigr) \mathcal{Y}_{l_{1}^{(1)}, s_{1}^{(1)}}^{J, m_{J}} (\bm{\hat{r}})
\end{multline}
where the effect of the short-range interaction amounts to the introduction of a phase shift $\eta_{J^{PC};1}^{(1)}$ with respect to the solution without mixing with $Q\bar{Q}$. Notice that here and in the following we omit bound components and focus instead on the asymptotic behavior of the open meson-meson components.

Let us now generalize to an arbitrary number of partial waves $(l_{k}^{(1)}, s_{k}^{(1)})$ coupling to $J^{PC}$ in the open channel. Recalling from the previous section that whenever there is more than one (partial-wave) channel there are many independent solutions with the same energy, the asymptotic behavior of any continuum solution can be conveniently expressed as
\begin{multline}
\psi^{(1)}_{J^{PC}, m_{J}; h}(\bm{r}) \simeq \frac{1}{r} \sqrt{\frac{2}{\pi} \frac{\mu^{(1)}}{p^{(1)}}} \sum_{k} i^{l_{k}^{(1)}} a_{J^{PC};k; h}^{(1)} \\
\sin \Bigl(p^{(1)} r - l_{k}^{(1)} \frac{\pi}{2} + \eta_{J^{PC};k; h}^{(1)} \Bigr) \mathcal{Y}_{l_{k}^{(1)}, s_{k}^{(1)}}^{J, m_{J}} (\bm{\hat{r}})
\end{multline}
(compare for example with Chap.~X, Eqs.~(12), (14) in \cite{Mot33}) where we have introduced the additional label $h$ to distinguish independent solutions with the same energy. Notice that, as the diabatic mixing couples the various meson-meson partial waves to each other, the coefficients $a_{J^{PC}; k; h}^{(1)}$ give account of the flow of probability from one partial-wave channel to the others through the coupled-channel interaction.

In the most general case of an arbitrary number of meson-meson components, one must realize that there is a different number of open channels depending on the energy. Then, if $E$ is the energy and $n$ the number of open thresholds at that energy, $E \ge T^{(j)}$ for $j=1,\dots, n$, the asymptotic behavior of the solutions is given by the open meson-meson components
\begin{multline}
\psi_{J^{PC}, m_{J}; h}^{(j)}(\bm{r}) \simeq \frac{1}{r} \sqrt{\frac{2}{\pi} \frac{\mu^{(j)}}{p^{(j)}}} \sum_{k} i^{l_{k}^{(j)}} a_{J^{PC}; k; h}^{(j)} \\
\sin\Bigr(p^{(j)} r - l^{(j)}_{k} \frac{\pi}{2} + \eta_{J^{PC}; k; h}^{(j)} \Bigl) \mathcal{Y}_{l_{k}^{(j)}, s_{k}^{(j)}}^{J, m_{J}}(\bm{\hat{r}})
\label{multasymcomp}
\end{multline}
where $h = 1, \dots, \tilde{n}$ with $\tilde{n}$ the total number of partial waves coupling to $J^{PC}$ in all the open meson-meson channels. For more than one open threshold, these solutions correspond to a meson-meson scattering process where the coupled-channel interaction is provided by the mixing of $Q\bar{Q}$ with all the meson-meson components.

\subsection{Meson-meson \texorpdfstring{$\mathrm{S}$}{S}-matrix}

Let us now see how to extract the on-shell meson-meson $\mathrm{S}$-matrix from the solutions of the diabatic \schr{} equation above threshold. As shown in the Appendix, the meson-meson scattering states from an initial open channel $j^{\prime}$, with quantum numbers $J^{PC}$, $m_{J}$, $l^{(j^{\prime})}$, $s^{(j^{\prime})}$, to a final open channel $j$ can be expressed as
\begin{multline}
\psi_{J^{PC}, m_{J};k^{\prime}}^{j \leftarrow j^{\prime}}(\bm{r}) \simeq \frac{1}{r} \sqrt{\frac{2}{\pi} \frac{\mu^{(j)}}{p^{(j)}}} \sum_{k} i^{l_{k}^{(j)}} \\
\Biggl[\delta_{j j^{\prime}} \delta_{k k^{\prime}} \sin\biggl(p^{(j)}r - l_{k}^{(j)}\frac{\pi}{2}\biggr) \\
+p^{(j)} f_{J^{PC}; k, k^{\prime}}^{j \leftarrow j^{\prime}} e^{i(p^{(j)} r - l_{k}^{(j)}\frac{\pi}{2})} \Biggr] \mathcal{Y}_{l_{k}^{(j)}, s_{k}^{(j)}}^{J, m_{J}}(\bm{\hat{r}})
\label{partscatt}
\end{multline}
where $k^{\prime}$ and $k$ are used to label distinct partial waves $(l^{(j^{\prime})}, s^{(j^{\prime})})$ and $(l^{(j)}, s^{(j)})$ coupling to $J^{PC}$ in the initial and final channel respectively, and $f_{J^{PC}; k, k^{\prime}}^{j \leftarrow j^{\prime}}$ is the corresponding partial-wave scattering amplitude. Notice that the normalization factors have been chosen according to the state normalization by energy
\begin{equation}
\braket{\Psi_{E} \vert \Psi_{E^{\prime}}} = \delta(E^{\prime} - E),
\end{equation}
instead of momentum. This is motivated by the fact that the open channels in general differ in their threshold and reduced mass, which makes normalization by momentum unpractical as there is a different value of the relative meson-meson momentum for each open channel.

It is important to realize that there are in total $\tilde{n}$ independent scattering states (one for each partial wave $k^{\prime}$ coupling to $J^{PC}$), that is as many as independent solutions to the diabatic \schr{} equation at the same energy. Hence Eqs.~\eqref{multasymcomp} and \eqref{partscatt} give two equivalent representations of the space of solutions at energy $E$, related to each other through a change of basis transformation. The change of basis transformation is written as
\begin{equation}
\psi_{J^{PC}, m_{J}; k^{\prime}}^{j \leftarrow j^{\prime}}(\bm{r}) = \sum_{h} \psi_{J, m_{J}; h}^{(j)}(\bm{r}) \Gamma_{J^{PC}; k^{\prime}; h}^{(j^{\prime})}
\label{chbase}
\end{equation}
where $\Gamma_{J^{PC}; k^{\prime}; h}^{(j^{\prime})}$ are the change of basis matrix elements. If we now plug \eqref{multasymcomp} and \eqref{partscatt} into \eqref{chbase}, after some simple algebra we obtain
\begin{equation}
\sum_{h} a_{J^{PC}; k; h}^{(j)} e^{-i \eta_{J^{PC}; k; h}^{(j)}} \Gamma_{J^{PC}; k^{\prime}; h}^{(j^{\prime})} =
\delta_{j j^{\prime}} \delta_{k k^{\prime}}
\label{rawrel}
\end{equation}
and
\begin{multline}
f_{J^{PC}; k, k^{\prime}}^{j \leftarrow j^{\prime}} = \frac{1}{2 i p^{(j)}} \\
\biggl(\sum_{h} a_{J^{PC}; k; h}^{(j)} e^{i \eta_{J^{PC}; k; h}^{(j)}} \Gamma_{J^{PC}; k^{\prime}; h}^{(j^{\prime})} - \delta_{j j^{\prime}} \delta_{k k^{\prime}} \biggr) .
\label{rawres}
\end{multline}

We can further simplify Eqs.~\eqref{rawrel} and \eqref{rawres} using matrix notation. For the sake of simplicity, let us rearrange momentarily the indices $(j, k)$, $(j^{\prime},  k^{\prime})$ as $\tilde{\jmath}, \tilde{\jmath}^{\prime}=1,\dots, \tilde{n}$, so that each $\tilde{\jmath}$ or $\tilde{\jmath}^{\prime}$ specifies uniquely a partial-wave channel for the scattering process. In this way we can drop the subscripts $k$, $k^{\prime}$ and rewrite Eqs.~\eqref{rawrel} and \eqref{rawres} as
\begin{equation}
\sum_{h} a_{J^{PC}; h}^{(\tilde{\jmath})} e^{-i \eta_{J^{PC}; h}^{(\tilde{\jmath})}} \Gamma_{J^{PC}; h}^{(\tilde{\jmath}^{\prime})} = \delta_{\tilde{\jmath} \tilde{\jmath}^{\prime}}
\label{simplirel}
\end{equation}
and
\begin{equation}
f_{J^{PC}}^{\tilde{\jmath} \leftarrow \tilde{\jmath}^{\prime}} = \frac{\sum_{h} a_{J^{PC}; h}^{(\tilde{\jmath})} e^{i \eta_{J^{PC}; h}^{(\tilde{\jmath})}} \Gamma_{J^{PC}; h}^{(\tilde{\jmath}^{\prime})} - \delta_{\tilde{\jmath} \tilde{\jmath}^{\prime}}}{2 i p^{(\tilde{\jmath})}}
\label{simplires}
\end{equation}
respectively. Let us now introduce the $\tilde{n}\times \tilde{n}$ Jost matrices $\mathscr{F}_{J^{PC}}^{\pm}$,
\begin{equation}
(\mathscr{F}^{\pm}_{J^{PC}})_{\tilde{\jmath} h} \equiv a_{J^{PC}; h}^{(\tilde{\jmath})} e^{\pm i \eta_{J^{PC}; h}^{(\tilde{\jmath})}}
\label{jost}
\end{equation}
where  $\tilde{\jmath}$ labels the rows and $h$ the columns, and the $\tilde{n}\times \tilde{n}$ change of basis matrix $\Gamma_{J^{PC}}$,
\begin{equation}
(\Gamma_{J^{PC}})_{h \tilde{\jmath}^{\prime}} \equiv \Gamma_{J^{PC}; h}^{(\tilde{\jmath}^{\prime})}
\end{equation}
with $h$ and $\tilde{\jmath}^{\prime}$ the row and column index respectively. Then Eq.~\eqref{simplirel} can be rewritten in matrix notation as
\begin{equation}
\mathscr{F}^{-}_{J^{PC}} \Gamma_{J^{PC}} = \mathbbold{1}
\end{equation}
where $\mathbbold{1}$ is the $\tilde{n}$-dimensional identity matrix, meaning that the change of basis matrix $\Gamma_{J^{PC}}$ is just the inverse of $\mathscr{F}^{-}_{J^{PC}}$. Plugging the change of basis matrix back into \eqref{simplires} yields the scattering amplitude
\begin{equation}
f^{\tilde{\jmath} \leftarrow \tilde{\jmath}^{\prime}}_{J^{PC}} = \frac{\bigl( \mathscr{F}^{+}_{J^{PC}}\bigr(\mathscr{F}^{-}_{J^{PC}}\bigl)^{-1} - \mathbbold{1} \bigr)_{\tilde{\jmath} \tilde{\jmath}^{\prime}}}{2 i p^{(\tilde{\jmath})}}
\label{theresult}
\end{equation}
from which, recalling that $J^{PC}$ conservation implies that the $\mathrm{S}$-matrix is block-diagonal
\begin{equation}
(\mathrm{S})_{J^{PC}, m_{J}, \tilde{\jmath}; J^{\prime P^{\prime} C^{\prime}}, m_{J^{\prime}}, \tilde{\jmath}^{\prime}} = \delta_{J^{PC} J^{\prime P^{\prime} C^{\prime}}} \delta_{m_{J} m_{J^{\prime}}} (\mathrm{S}_{J^{PC}})_{\tilde{\jmath} \tilde{\jmath}^{\prime}},
\end{equation}
one can recognize the $J^{PC}$ block of the $\mathrm{S}$-matrix as
\begin{equation}
\mathrm{S}_{J^{PC}} = \mathscr{F}^{+}_{J^{PC}} \bigl(\mathscr{F}^{-}_{J^{PC}}\bigr)^{-1}.
\label{smadef}
\end{equation}

Equation~\eqref{smadef} is a general formula known in multichannel scattering theory, see for example Eq.~(20.18) in \cite{Tay72} (notice that we lack the momentum factors because we adopt a different normalization), that allows us to calculate the on-shell $\mathrm{S}$-matrix numerically from the solution of the diabatic \schr{} equation above threshold. Indeed, the numerical values of the amplitudes $a_{J^{PC}; k; h}^{(j)}$ and phase shifts $\eta_{J^{PC}; k; h}^{(j)}$, defining the Jost matrices \eqref{jost}, can be obtained simply by fitting Eq.~\eqref{multasymcomp} to the long-distance numerical wave functions from Sec.~\ref{numsec}.

Finally, it is shown in the Appendix that the total unpolarized cross-section can be calculated as
\begin{equation}
\sigma^{j \leftarrow j^{\prime}} = \sum_{J^{PC}} \sigma^{j \leftarrow j^{\prime}}_{J^{PC}}
\label{totcsec}
\end{equation}
with each $J^{PC}$ cross section $\sigma^{j \leftarrow j^{\prime}}_{J^{PC}}$ obtained as
\begin{equation}
\sigma^{j \leftarrow j^{\prime}}_{J^{PC}} = \frac{4 \pi (2 J + 1)}{(2 s_{M_{1}^{(j^{\prime})}} + 1) (2 s_{\bar{M}_{2}^{(j^{\prime})}} + 1)} \sum_{k} \sum_{k^{\prime}} \bigl\lvert f_{J^{PC}; k, k^{\prime}}^{j \leftarrow j^{\prime}}\bigr\rvert^{2}
\label{partcsec}
\end{equation}
where $s_{M_{1}^{(j^{\prime})}}$ and $s_{\bar{M}_{2}^{(j^{\prime})}}$ are the spins of the initial-channel mesons $M_{1}^{(j^{\prime})}$ and $\bar{M}_{2}^{(j^{\prime})}$, and we have restored the notation $(j, k)$, $(j^{\prime},  k^{\prime})$ for the partial-wave channels.

\section{Results\label{ressec}}

In this section we identify quarkoniumlike mesons from the structures in the calculated $J^{PC}$ cross sections as a function of the energy. More precisely, we do not use the total cross section \eqref{totcsec} but rather a scaled $J^{PC}$ cross section defined as
\begin{equation}
\begin{split}
\bar{\sigma}^{j \leftarrow j^{\prime}}_{J^{PC}} &\equiv \frac{(2 s_{M_{1}^{(j^{\prime})}} + 1) (2 s_{\bar{M}_{2}^{(j^{\prime})}} + 1)}{4 \pi (2 J + 1)} (p^{(j)})^2 \sigma^{j \leftarrow j^{\prime}}_{J^{PC}} \\
&= \sum_{k} \sum_{k^{\prime}} \bigl\lvert p^{(j)} f_{J^{PC}; k, k^{\prime}}^{j \leftarrow j^{\prime}}\bigr\rvert^{2}.
\end{split}
\label{scaledcs}
\end{equation}
This scaled $J^{PC}$ cross section is more convenient for our theoretical analysis for a number of reasons:
\begin{itemize}
\item it makes easier the distinction of resonances;
\item it is a dimensionless quantity;
\item it is not affected by the purely kinematical $(p^{(j)})^{-2}$ behavior of the cross section, what allows for a more detailed study for energies close above threshold;
\item it is symmetric under exchanges $j \leftrightarrow j^{\prime}$, as per the detailed balance principle (i.e., time reversal symmetry);
\item its values are bounded by the unitarity condition $\sum_{j, j^{\prime}} \bar{\sigma}^{j \leftarrow j^{\prime}}_{J^{PC}} \le 1$, where the maximum value of $1$ is expected at the mass of an isolated resonance with little nonresonant background.
\end{itemize}
We restrict our study to meson-meson pairs with pretty small widths and thresholds well separated in energy. In this manner we avoid the technical complications deriving from the treatment of the widths and the possible presence of overlaping thresholds. This constrains our formalism to be safely applicable only up to energies of $4.1$~GeV in the charmoniumlike sector, see \cite{Bru20,Bru21}, and $10.8$~GeV in the bottomoniumlike one, see \cite{DiaBot}. For the sake of simplicity we shall discuss here only elastic processes, as they are the most convenient for the current theoretical analysis.

For practical purposes we shall compare the scattering resonances with existing quarkoniumlike states and with our former predictions from the bound state approximation. For this comparison to be meaningful we use the same parameters as in Refs.~\cite{Bru20,Bru21,DiaBot}, that we list here for completeness.

For the Cornell potential we take the standard
\begin{align}
\sigma &  =925.6\text{~MeV/fm},\\
\chi &  =102.6\text{~MeV~fm},
\end{align}
(we use $\hbar=c=1$ units, so $1\text{~fm}^{-1} \approx 197.3\text{~MeV}$) while for the constant we choose a flavor-independent value
\begin{equation}
\beta = 855\text{~MeV},
\end{equation}
and for the heavy quark masses we use
\begin{align}
m_{c} &  =1840\text{~MeV},\\
m_{b} &  =5215\text{~MeV}.
\end{align} 

As for the mixing potential \eqref{mixpot}, in order not to spoil our predictive power we choose universal values $\rho$ and $\Delta$ for all thresholds. Moreover, we adopt the prescription of taking $\Delta$ as the effective mixing strength of $Q\bar{Q}$ with the isospin-singlet combination of two approximately degenerate thresholds containing up and down quarks. Then the effective mixing strength of $Q\bar{Q}$ with a hidden-strange threshold is given by $\Delta / \sqrt{2}$ (see Appendix in Ref.~\cite{DiaBot} for more details). We use
\begin{align}
\Delta_{c} &= 130\text{~MeV}, \\
\rho_{c} &= 0.3\text{~fm},
\end{align}
for charmoniumlike systems and
\begin{align}
\Delta_{b} &= 55\text{~MeV}, \\
\rho_{b} &= 0.3\text{~fm}.
\end{align}
for bottomoniumlike ones.

Finally, the threshold values have been obtained from the sum of the corresponding open-flavor meson experimental masses \cite{PDG20}.

The calculated scaled $J^{PC}=(0, 1, 2)^{++}$ and $J^{PC}=1^{--}$ cross sections for the elastic scattering of open-charm and open-bottom meson-meson pairs are presented in Subsec.~\ref{charm} and \ref{bottom} respectively. It may be useful to anticipate their common general features:
\begin{enumerate}[i)]
\item
There is a resonance at about the energy of any Cornell quarkonium state. These resonances, containing a very dominant $Q\bar{Q}$ component, are usually most evident in the open channel with the nearest threshold.
\item
There appear additional resonances that do not correspond to the energy of any Cornell state. These resonances, with a significant meson-meson component, are located close to some meson-meson threshold coupling to $J^{PC}$.
\item
Superposition of different resonances and their overlap with thresholds may result in the creation of complex structures in the cross section which deviate from the customary Breit-Wigner form.
\end{enumerate}
The emerging physical picture is that of a quarkoniumlike spectrum consisting in quasi-conventional plus unconventional states from the meson-meson interaction provided by the diabatic mixing of Fig.~\ref{qqexch}. The appearance of the unconventional states, as well as their eventual composition, depends on the strength of the $Q\bar{Q}$--meson-meson mixing and the position of the Cornell potential bound states with respect to the meson-meson thresholds coupling to $J^{PC}$.

In what follows we simply identify the enhancements in the calculated open-charm and open-bottom cross sections which can be associated to resonances. A more thorough analysis, especially of the most complex structures such as threshold cusps and minimums, shall be the subject of a forthcoming paper.

\subsection{Elastic cross section for open-charm mesons\label{charm}}

\begin{figure*}
\input{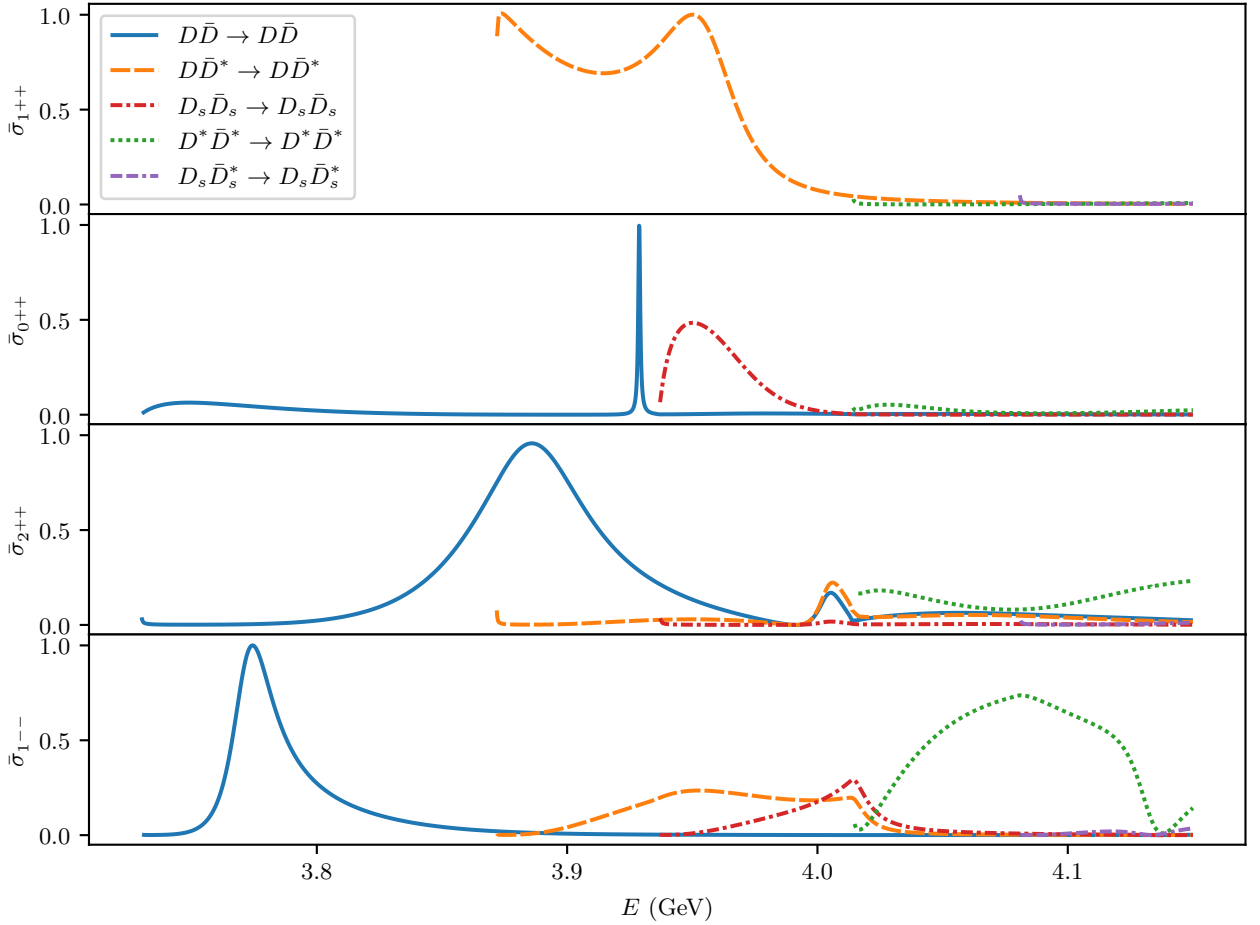}
\caption{Scaled elastic cross section $\bar{\sigma}_{J^{PC}}$ for open-charm meson-meson scattering with $J^{PC}=(0,1,2)^{++}$ and $J^{PC}=1^{--}$, versus c.m.\ energy $E$ in GeV.\label{charm_cs}}
\end{figure*}

The calculated scaled cross section for elastic open-charm meson-meson scattering processes with $J^{PC}= (0,1,2)^{++}$ and $J^{PC} = 1^{--}$ are shown in Fig.~\ref{charm_cs}.

Visual comparison of these plots with the masses and widths reported in Table~V of Ref.~\cite{Bru21} provides us with an independent check of the results
obtained in the bound state approximation. So the presence of a narrow resonance with $J^{PC}=0^{++}$ close below the $D_{s}\bar{D}_{s}$ threshold with a mass of $3920.9$~MeV, coming from the interaction of the $D_{s}\bar{D}_{s}$ threshold with the $2P$ Cornell state, is confirmed from the calculated peak in the $D\bar{D}$ cross section. Another peak, without correspondence in the bound state approximation, is visible in the $D_{s}\bar{D}_{s}$ cross section close above threshold, which may include some enhancement due to the presence of the resonance close below threshold. This second peak at about $3950$~MeV can be partly attributed to a quasi-conventional $0^{++}$ resonance with a very dominant $2P$ $c\bar{c}$ component coming as well from the interaction of the $2P$ Cornell state at $3953.7$~MeV with the $D_{s}\bar{D}_{s}$ threshold. Notice that this state was also present in the bound state approximation when neglecting the (open) $D_{s}\bar{D}_{s}$ threshold, but it was discarded because a one-to-one correspondence with Cornell states was assumed and the $2P$ $c\bar{c}$ core had already been assigned to the calculated bound state at $3920.9$~MeV \cite{Bru21}. These two peaks may be related to the experimental candidates $X(3915)$ (see below) and the more uncertain $\chi_{c0}(3860)$. Indeed, it is quite possible that due to the proximity in energy of the peaks their effect is present in both candidates. If this is the case, dedicated experiments would be required to unveil their presence. In this regard it is important to emphasize that these peaks are more easily distinguished in the scaled cross section \eqref{scaledcs} than in the physical one \eqref{partcsec}.

Apart from these peaks a tiny bump at about $4030$~MeV can be seen in the $D^{\ast}\bar{D}^{\ast}$ cross section. It might correspond to a new resonance due to the interaction between the $D^{\ast}\bar{D}^{\ast}$ threshold and the distant $2P$ Cornell state, but its experimental observation in the physical cross section seems unlikely. (Even more difficult should be the observation of the other tiny structure at the $D\bar{D}$ threshold.)

For $J^{PC}=2^{++}$ the calculated $D\bar{D}$ elastic cross section shows two peaks, a broader one at about $3.9$~GeV, close above the $D\bar{D}^{\ast}$ threshold, and a narrower one at about $4$~GeV, close below the $D^{\ast}\bar{D}^{\ast}$ threshold. They correspond to the two excited $2^{++}$ resonances at $3881.1$~MeV and $4003.9$~MeV calculated in the bound state approximation, the broader (narrower) one  resulting mainly from the threshold interactions with the $2P$ $(1F)$ Cornell state. Notice that the narrower peak around $4$~GeV is also visible in the $D\bar{D}^{\ast}$ and, to a much lesser extent, in the $D_{s}\bar{D}_{s}$ cross sections. From the experimental point of view there is a $2^{++}$ candidate, the $\chi_{c2}(3930)$, which may be assigned to the broader peak.

It may be worth mentioning that there is a tiny enhancement of the $D\bar{D}^{\ast}$ cross section on the background of the broader peak, near $3.95$~GeV, that can be associated to a quasi-conventional resonance with a very dominant $2P$ $c\bar{c}$ component. Moreover, a quasi-conventional resonance with a very dominant $1F$ $c\bar{c}$ component can be inferred from the bump in the $D^{\ast} \bar{D}^{\ast}$ cross section just above threshold. Neither of these resonances find correspondence in the bound state approximation since the $2P$ and $1F$ $c\bar{c}$ cores were assigned to the bound states corresponding to the peaks near $3.9$~GeV and $4$~GeV respectively.

For $J^{PC}=1^{++}$ two peaks are clearly visible in the $D\bar{D}^{\ast}$ cross section, one very close to threshold, which can be correlated to the presence of the $\chi_{c1}(3872)$ state close below threshold, and another one at about $3.95$~GeV, corresponding to a quasi-conventional $2P$ $c\bar{c}$ resonance, that has no correspondence in the bound state approximation. This last state was discarded in Ref.~\cite{Bru21} because the $2P$ $c\bar{c}$ core had already been assigned to the calculated bound state at $3871.7$~MeV. At present, there is no suitable experimental candidate in the PDG \cite{PDG20} for this $1^{++}$ resonance at about $3.95$~GeV. We note however that recent studies of $e^{+}e^{-}\to \gamma \, \omega J\!/\!\psi$ from the BESIII Collaboration \cite{Abl19} show that a good fit to data is obtained by introducing either two or three resonant structures. In the three-resonance fitting scenario, along with the $X(3872)$ and $X(3915)$ there is an additional resonance, labeled $X(3960)$, at $3963.7\pm5.5$~MeV (see Table~I of that reference). Then our calculated $1^{++}$ peak at $3.95$~GeV may be tentatively identified with $X(3960)$. Moreover, notice that according to \cite{Abl19} the fitting including $X(3872)$, $X(3915)$, and $X(3960)$ implies a width for $X(3915)$ much smaller than the current PDG average value, opening an alternative interpretation of this state as the experimental counterpart of the very narrow $0^{++}$ calculated peak at $3920.9$~MeV (see above). We strongly encourage further experimental efforts in establishing the possible existence of a $1^{++}$ charmoniumlike resonance near $3.95$~GeV as well as the quantum numbers of the $X(3915)$, what could provide a crucial test of our predictions.

For $J^{PC}=1^{--}$ a quasi-conventional resonance around $3.77$~GeV is clearly visible in the calculated $D\bar{D}$ cross section, in perfect correspondence with the state of mass $3771.7$~MeV (with a very dominant $1D$ $c\bar{c}$ component) obtained in the bound state approximation and with the experimental $\psi(3770)$. On the other hand the $D^{\ast}\bar{D}^{\ast}$ cross section shows a peak at $4080$~MeV, close below the $D_{s}\bar{D}_{s}^{\ast}$ threshold, which may be assigned to the experimental $\psi(4040)$. This is particularly relevant because, as explained in Ref.~\cite{Bru21}, the bound state approximation is not suited for the description of this resonance. In this regard it is worth mentioning that the calculated bump is not a standard Breit-Wigner curve. This complex structure seems to include the effect of a quasi-conventional $3S$ $c\bar{c}$ resonance located around $4097$~MeV, and maybe also some effect from a third resonance above $4.1$~GeV (as an extended numerical calculation beyond $4.1$~GeV seems to point out). The minimum around $4.14$~GeV in the calculated cross sections has to do with this complexity.

In addition to the aforementioned minimum, the calculated open-charm meson-meson scaled cross sections reveal the presence of other peculiar structures such as the threshold cusps in the $1^{--}$ $D_{s}\bar{D}_{s}$ and $D\bar{D}^{\ast}$ cross sections at the $D^{\ast}\bar{D}^{\ast}$ threshold. Threshold cusps have been studied extensively in the literature, see for example \cite{Guo20} and references therein, but it is not clear \textit{a priori} which thresholds may produce a nontrivial structure. In this respect, the diabatic formalism may be an ideal framework to study their occurrence as it provides a unified description for energies across various thresholds. An analysis of the calculated threshold cusps in the charm as well as in the bottom sectors will be given in a separated paper.

\subsection{Elastic cross section for open-bottom mesons\label{bottom}}

\begin{figure*}
\input{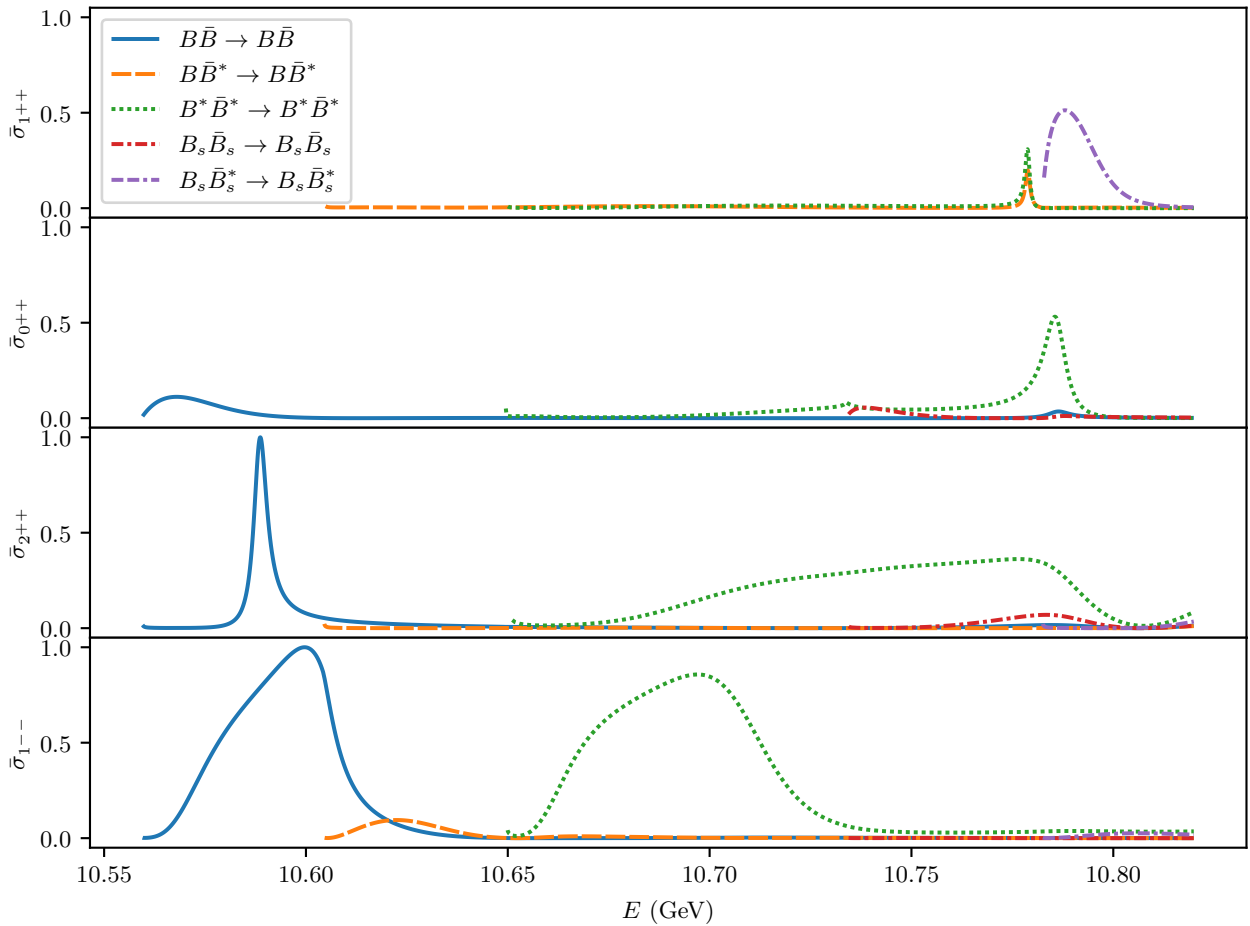}
\caption{Scaled elastic cross section $\bar{\sigma}_{J^{PC}}$ for open-bottom meson-meson scattering with $J^{PC}=(0,1,2)^{++}$ and $J^{PC}=1^{--}$, versus c.m.\ energy $E$ in GeV.\label{bottom_cs}}
\end{figure*}

The calculated scaled cross section for elastic open-bottom meson-meson scattering processes with $J^{PC}=(0,1,2)^{++}$ and $J^{PC}=1^{--}$ are shown in Fig.~\ref{bottom_cs}.

As in the charmoniumlike case, the calculated cross section confirm the bottomoniumlike resonances calculated within the bound state approximation, see Table~VIII of Ref.~\cite{DiaBot}. We see that for $0^{++}$ there is a peak at about the energy of the $4P$ Cornell state at $10782.2$~MeV, corresponding to a quasi-conventional $4P$ $b\bar{b}$ resonance (recall that the nearby $B_{s}\bar{B}_{s}^{\ast}$ threshold does not couple to $0^{++})$. This peak, much more visible in the $B^{\ast}\bar{B}^{\ast}$ than in the $B\bar{B}$ and $B_{s}\bar{B}_{s}$ cross sections, corresponds to the bound state predicted at $10778.1$~MeV. Although no experimental candidate is known at present, it has to be remarked that this resonance shows up as a standard Breit-Wigner peak since it is quite isolated from any other structure. This may facilitate the experimental discovery of the corresponding bottomiumlike resonance, notwithstanding its relatively small width. Actually, apart from this peak there is only a tiny cusp in the $B^{\ast}\bar{B}^{\ast}$ cross section at the $B_{s}\bar{B}_{s}$ threshold and a small bump in $B\bar{B}$ just above threshold which may be due to the interaction of $B\bar{B}$ with the $3P$ Cornell state at $10536.6$~MeV.

For $2^{++}$ there are two relevant structures: a Breit-Wigner peak in the $B\bar{B}$ cross section, somewhat below the $B\bar{B}^{\ast}$ threshold, and a 
very broad and asymmetric bump (most visible in the $B^{\ast} \bar{B}^{\ast}$ cross section) reaching its maximum around the $B_{s} \bar{B}_{s}^{\ast}$ threshold. The first peak corresponds to a quasi-conventional resonance, with a very dominant $2F$ $b\bar{b}$ component, predicted in the bound state approximation (the effect of the $B\bar{B}^{\ast}$ threshold being just a small shift in the mass of the $2F$ $b\bar{b}$ Cornell state). On the other hand, the second more complex structure may be involving the effect of more than one resonance: the one at $10782.3$~MeV from the bound state approximation and some other resonance above $10.8$~GeV, as the presence of a minimum might be indicating (and an extended numerical calculation beyond $10.8$~GeV seems to confirm). It may also include some effect from the interaction of the $B^{\ast}\bar{B}^{\ast}$ threshold with the $2F$ Cornell state at $10601.0$~MeV. This complexity may make difficult the identification of the underlying resonance(s). In contrast, we expect the Breit-Wigner peak in $B\bar{B}$ close below the $B\bar{B}^{\ast}$ threshold to be clearly identified when data are available. 

For $1^{++}$ there is a peak close below the $B_{s}\bar{B}_{s}^{\ast}$ threshold that is clearly visible in the $B^{\ast}\bar{B}^{\ast}$ and $B\bar{B}^{\ast}$ cross sections, corresponding to the resonance at $10778.9$~MeV from the bound state approach. In addition to this peak there appears an enhancement in the $B_{s}\bar{B}_{s}^{\ast}$ cross section just above the corresponding threshold, at about the mass of the $4P$ Cornell state ($10782.2$~MeV). The presence of the quasi-conventional $4P$ $b\bar{b}$ resonance in this structure is somehow diluted within the enhancement of the $B_{s} \bar{B}_{s}^{\ast}$ cross section due to the presence of the resonance close below threshold. In \cite{DiaBot} we suggested to look experimentally for a narrow resonance at about $10780$~MeV, which is in perfect agreement with the sharp peak calculated just below the $B_{s}\bar{B}_{s}^{\ast}$ threshold. We realize now that the presence of the second peak just above the $B_{s}\bar{B}_{s}^{\ast}$ threshold could hinder its experimental detection. In this regard, the expected decay of the narrow resonance to $\Upsilon(1S)\phi$ could be determinant for its disentanglement from data.

For $1^{--}$ the two prominent peaks observed in the figure are in perfect agreement with the bound state approximation prediction. The first peak, close below the $B\bar{B}^{\ast}$ threshold, comes mostly from the interaction of the $B\bar{B}^{\ast}$ threshold with the $4S$ $b\bar{b}$ state at $10615.0$~MeV. Also related to this interaction there is a bump in the $B\bar{B}^{\ast}$ cross section close above threshold which is associated to a quasi-conventional $4S$ $b\bar{b}$ resonance. As for the second prominent peak it can be associated to a quasi-conventional $3D$ $b\bar{b}$ resonance. Experimentally, the prominent peaks can be assigned to the well established $\Upsilon(10580)$ and the not well established $\Upsilon(10753)$ respectively, whereas the effect of the smaller bump is expected to be diluted within the data assigned to $\Upsilon(10580)$. Notice that both prominent peaks are slightly asymmetric, possibly indicating additional contributions from a threshold interaction with the $3D$ Cornell state in the first peak and with the $4S$ Cornell state in the second one.

Altogether, the results for charmoniumlike and bottomoniumlike states show, as anticipated, a spectrum of quasi-conventional resonances in one-to-one correspondence with the Cornell spectrum plus additional unconventional resonances located close to some meson-meson thresholds.

It is worth remarking that these results, which overcome the shortcomings of the bound state approximation, allow for a description of all existing candidates in the energy range under study as well as for definite prescriptions which could guide the experimental searches of yet unknown predicted resonances.

\section{Summary\label{endsec}}

The diabatic framework, a QCD-based formalism for the description of quarkoniumlike systems in terms of $Q\bar{Q}$ and open flavor meson-meson components, has been extended to the study of coupled-channel meson-meson scattering. This has allowed to overcome the shortcomings of the bound state approximation previously used for the analysis of quarkoniumlike states with masses above the lowest meson-meson threshold.

Our study of coupled-channel meson-meson scattering has gone through two separate steps. First, we have solved numerically the diabatic \schr{} equation above threshold by using an extension of the well known grid method which allows for nonvanishing values of the wave function at the boundaries. Then, we have used the asymptotic behavior of these continuum solutions to extract the meson-meson scattering amplitudes and cross sections.

This procedure has been applied to the calculation of the elastic cross section as a function of the energy for open-charm as well as open-bottom meson-meson scattering. Quarkoniumlike resonances have been identified from standard Breit-Wigner peaks and more general structures in the cross section. We have shown that quarkoniumlike resonances previously obtained from the bound state approximation constitute only a part (as a direct consequence of the constraints inherent to such approximation) of the whole spectrum. We find that this is formed by quasi-conventional resonances, with masses close to those of Cornell bound states, as well as by unconventional  ones, with masses close to the energies of some meson-meson thresholds. It must be observed that sometimes the quasi-conventional resonances are not isolated from other enhancements in the cross section, which hinders their identification. All existing experimental candidates have been assigned to calculated states. As for predicted but yet undiscovered resonances, an initial analysis to guide experimental searches has been outlined. In this regard, the experimental confirmation of our predictions of a $1^{++}$ quasi-conventional charmoniumlike resonance with a mass about $3950$~MeV, a narrow $0^{++}$ unconventional charmoniumlike resonance with a mass about $3920$~MeV assigned to $X(3915)$, and a $1^{++}$ unconventional bottomoniumlike resonance with a mass about $10780$~MeV, would provide strong further support to our theoretical description.

\begin{acknowledgments}
This work has been supported by \foreignlanguage{spanish}{Ministerio de Ciencia e Innovaci\'on} and \foreignlanguage{spanish}{Agencia Estatal de Investigaci\'on} of Spain and European Regional Development Fund Grant PID2019-105439 GB-C21 and by EU Horizon 2020 Grant No.~824093 (STRONG-2020). R.~B. acknowledges a FPI fellowship from \foreignlanguage{spanish}{Ministerio de Ciencia, Innovaci\'on y Universidades} of Spain under Grant No. BES-2017-079860.
\end{acknowledgments}

\onecolumngrid

\appendix*

\section{Partial-wave expansion of meson-meson scattering states}

Since we are treating heavy-light quark mesons as pointlike particles, the stationary scattering state from an initial meson-meson channel $M_{1}^{(j^{\prime})} \bar{M}_{2}^{(j^{\prime})}$ with spins $s_{M_{1}^{(j^{\prime})}}$, $s_{\bar{M}_{2}^{(j^{\prime})}}$ and projections $\sigma_{M_{1}^{(j^{\prime})}}$, $\sigma_{\bar{M}_{2}^{(j^{\prime})}}$ to a final channel $M_{1}^{(j)} \bar{M}_{2}^{(j)}$ with spins $s_{M_{1}^{(j)}}$, $s_{\bar{M}_{2}^{(j)}}$ and projections $\sigma_{M_{1}^{(j)}}$, $\sigma_{\bar{M}_{2}^{(j)}}$ can be written as (see for example Eq.~(20.11) in \cite{Tay72})
\begin{equation}
\psi_{\bm{\hat{p}}; \sigma_{M_{1}^{(j)}}, \sigma_{\bar{M}_{2}^{(j)}}; \sigma_{M_{1}^{(j^{\prime})}}, \sigma_{\bar{M}_{2}^{(j^{\prime})}}}^{j \leftarrow j^{\prime}}(\bm{r}) \simeq \frac{\sqrt{\mu^{(j)} p^{(j)}}}{(2 \pi)^{\frac{3}{2}}}
\Biggl( \delta_{j j^{\prime}} \delta_{\sigma_{M_{1}^{(j)}} \sigma_{M_{1}^{(j^{\prime})}}} \delta_{\sigma_{\bar{M}_{2}^{(j)}} \sigma_{\bar{M}_{2}^{(j^{\prime})}}} e^{i \bm{p}^{(j^{\prime})} \cdot \bm{r}}
+ f^{j \leftarrow j^{\prime}}_{\sigma_{M_{1}^{(j)}}, \sigma_{\bar{M}_{2}^{(j)}}; \sigma_{M_{1}^{(j^{\prime})}}, \sigma_{\bar{M}_{2}^{(j^{\prime})}}}(\bm{\hat{p}} \cdot \bm{\hat{r}}) \frac{e^{i p^{(j)} \, r}}{r} \Biggr)
\label{scadef}
\end{equation}
where $\bm{\hat{p}}$ and $\bm{\hat{r}}$ are interpreted as the beam and detection directions respectively, with $f^{j \leftarrow j^{\prime}}_{\sigma_{M_{1}^{(j)}}, \sigma_{\bar{M}_{2}^{(j)}}; \sigma_{M_{1}^{(j^{\prime})}}, \sigma_{\bar{M}_{2}^{(j^{\prime})}}}(\bm{\hat{p}} \cdot \bm{\hat{r}})$ the corresponding scattering amplitude. Notice that for the sake of simplicity we have omitted the subscripts $s_{M_{1}^{(j)}}$, $s_{\bar{M}_{2}^{(j)}}$, $s_{M_{1}^{(j^{\prime})}}$, $s_{\bar{M}_{2}^{(j^{\prime})}}$ in the scattering states, in the understanding that they are always implicit.

We wish to expand \eqref{scadef} in stationary scattering states with definite $J$, $m_{J}$, $l^{(j^{\prime})}$, and $s^{(j^{\prime})}$ quantum numbers. To do so, we introduce the plane wave expansion
\begin{multline}
\delta_{j j^{\prime}} \delta_{\sigma_{M_{1}^{(j)}} \sigma_{M_{1}^{(j^{\prime})}}} \delta_{\sigma_{\bar{M}_{2}^{(j)}} \sigma_{\bar{M}_{2}^{(j^{\prime})}}} e^{i \bm{p}^{(j^{\prime})} \cdot \bm{r}} = \\ 4 \pi \sum_{l^{(j)}, m_{l^{(j)}}} \sum_{l^{(j^{\prime})}, m_{l^{(j^{\prime})}}} \delta_{j j^{\prime}} \delta_{l^{(j)} l^{(j^{\prime})}} \delta_{m_{l^{(j)}} m_{l^{(j^{\prime})}}} \delta_{\sigma_{M_{1}^{(j)}} \sigma_{M_{1}^{(j^{\prime})}}} \delta_{\sigma_{\bar{M}_{2}^{(j)}} \sigma_{\bar{M}_{2}^{(j^{\prime})}}} i^{l^{(j)}} j_{l^{(j)}}(p^{(j)} r) Y_{l^{(j)}}^{m_{l^{(j)}}}(\bm{\hat{r}}) Y_{l^{(j^{\prime})}}^{m_{l^{(j^{\prime})}} \ast} (\bm{\hat{p}})
\end{multline}
and rearrange it conveniently as
\begin{multline}
\delta_{j j^{\prime}} \delta_{\sigma_{M_{1}^{(j)}} \sigma_{M_{1}^{(j^{\prime})}}} \delta_{\sigma_{\bar{M}_{2}^{(j)}} \sigma_{\bar{M}_{2}^{(j^{\prime})}}} e^{i \bm{p}^{(j^{\prime})} \cdot \bm{r}} =
4 \pi \sum_{J, m_{J}} \sum_{l^{(j)}, m_{l^{(j)}}} \sum_{s^{(j)}, m_{s^{(j)}}} \sum_{l^{(j^{\prime})}, m_{l^{(j^{\prime})}}} \sum_{s^{(j^{\prime})}, m_{s^{(j^{\prime})}}} \delta_{j j^{\prime}} \delta_{l^{(j)} l^{(j^{\prime})}} \delta_{s^{(j)} s^{(j^{\prime})}} i^{l^{(j)}} j_{l^{(j)}}(p^{(j)} r) \\
C_{s_{M_{1}^{(j)}}, s_{\bar{M}_{2}^{(j)}}, s^{(j)}}^{\sigma_{M_{1}^{(j)}}, \sigma_{\bar{M}_{2}^{(j)}}, m_{s^{(j)}}} C_{l^{(j)}, s^{(j)}, J}^{m_{l^{(j)}}, m_{s^{(j)}}, m_{J}}
Y_{l^{(j)}}^{m_{l^{(j)}}}(\bm{\hat{r}})
C_{s_{M_{1}^{(j^{\prime})}}, s_{\bar{M}_{2}^{(j^{\prime})}}, s^{(j^{\prime})}}^{\sigma_{M_{1}^{(j^{\prime})}}, \sigma_{\bar{M}_{2}^{(j^{\prime})}}, m_{s^{(j^{\prime})}}} C_{l^{(j^{\prime})}, s^{(j^{\prime})}, J}^{m_{l^{(j^{\prime})}}, m_{s^{(j^{\prime})}}, m_{J}}
Y_{l^{(j^{\prime})}}^{m_{l^{(j^{\prime})}} \ast}(\bm{\hat{p}}).
\label{plaexp}
\end{multline}
Similarly, we expand the scattering amplitude as
\begin{equation}
f^{j \leftarrow j^{\prime}}_{\sigma_{M_{1}^{(j)}}, \sigma_{\bar{M}_{2}^{(j)}}; \sigma_{M_{1}^{(j^{\prime})}}, \sigma_{\bar{M}_{2}^{(j^{\prime})}}}(\bm{\hat{p}} \cdot \bm{\hat{r}}) = 4 \pi \sum_{l^{(j)}, m_{l^{(j)}}} \sum_{l^{(j^{\prime})}, m_{l^{(j^{\prime})}}} 
f_{l^{(j)}, m_{l^{(j)}}; l^{(j^{\prime})}, m_{l^{(j^{\prime})}}; \sigma_{M_{1}^{(j)}}, \sigma_{\bar{M}_{2}^{(j)}}; \sigma_{M_{1}^{(j^{\prime})}}, \sigma_{\bar{M}_{2}^{(j^{\prime})}}}^{j \leftarrow j^{\prime}}
Y_{l^{(j)}}^{m_{l^{(j)}}}(\bm{\hat{r}}) Y_{l^{(j^{\prime})}}^{m_{l^{(j^{\prime})}} \ast}(\bm{\hat{p}}),
\end{equation}
which using angular momentum algebra can be rewritten as
\begin{multline}
f^{j \leftarrow j^{\prime}}_{\sigma_{M_{1}^{(j)}}, \sigma_{\bar{M}_{2}^{(j)}}; \sigma_{M_{1}^{(j^{\prime})}}, \sigma_{\bar{M}_{2}^{(j^{\prime})}}}(\bm{\hat{p}} \cdot \bm{\hat{r}}) =
4 \pi \sum_{J, m_{J}} \sum_{l^{(j)}, m_{l^{(j)}}} \sum_{s^{(j)}, m_{s^{(j)}}} \sum_{l^{(j^{\prime})}, m_{l^{(j^{\prime})}}} \sum_{s^{(j^{\prime})}, m_{s^{(j^{\prime})}}} f_{J; l^{(j)},s^{(j)}; l^{(j^{\prime})},s^{(j^{\prime})}}^{j \leftarrow j^{\prime}} \\
C_{s_{M_{1}^{(j)}}, s_{\bar{M}_{2}^{(j)}}, s^{(j)}}^{\sigma_{M_{1}^{(j)}}, \sigma_{\bar{M}_{2}^{(j)}}, m_{s^{(j)}}} C_{l^{(j)}, s^{(j)}, J}^{m_{l^{(j)}}, m_{s^{(j)}}, m_{J}}
Y_{l^{(j)}}^{m_{l^{(j)}}}(\bm{\hat{r}})
C_{s_{M_{1}^{(j^{\prime})}}, s_{\bar{M}_{2}^{(j^{\prime})}}, s^{(j^{\prime})}}^{\sigma_{M_{1}^{(j^{\prime})}}, \sigma_{\bar{M}_{2}^{(j^{\prime})}}, m_{s^{(j^{\prime})}}} C_{l^{(j^{\prime})}, s^{(j^{\prime})}, J}^{m_{l^{(j^{\prime})}}, m_{s^{(j^{\prime})}}, m_{J}}
Y_{l^{(j^{\prime})}}^{m_{l^{(j^{\prime})}} \ast}(\bm{\hat{p}}).
\label{ampexp}
\end{multline}
Let us now take the definition of the meson-meson spin-momentum eigenfunctions
\begin{equation}
\mathcal{Y}_{l, s}^{J, m_{J}}(\bm{\hat{x}}) = \sum_{m_{l}, m_{s}}
C_{l, s, J}^{m_{l}, m_{s}, m_{J}} Y_{l}^{m_{l}}(\bm{\hat{x}}) \sum_{\sigma^{\prime}_{M_{1}}, \sigma^{\prime}_{\bar{M}_{2}}} C_{s_{M_{1}}, s_{\bar{M}_{2}}, s}^{\sigma^{\prime}_{M_{1}}, \sigma^{\prime}_{\bar{M}_{2}}, m_{s}} \xi_{s_{M_1}}^{\sigma^{\prime}_{M_1}} \xi_{s_{\bar{M}_2}}^{\sigma^{\prime}_{\bar{M}_2}}
\end{equation}
with $\xi_{s_{M_1}}^{\sigma^{\prime}_{M_1}}$ and $\xi_{s_{\bar{M}_2}}^{\sigma^{\prime}_{\bar{M}_2}}$ the meson spin vectors, where $\bm{\hat{x}}$ can be either $\bm{\hat{r}}$ or $\bm{\hat{p}}$, $l$ either $l^{(j)}$ or $l^{(j^{\prime})}$, $s$ either $s^{(j)}$ or $s^{(j^{\prime})}$, $M_1$ either $M_1^{(j)}$ or $M_1^{(j^{\prime})}$, and $\bar{M}_2$ either $\bar{M}_2^{(j)}$ or $\bar{M}_2^{(j^{\prime})}$. Using orthogonality of the meson spin vectors, $\xi_{s_{M_1}}^{\sigma_{M_1} \dagger} \xi_{s_{M_1}}^{\sigma^{\prime}_{M_1}} = \delta_{\sigma_{M_1} \sigma^{\prime}_{M_1}}$ and $\xi_{s_{\bar{M}_2}}^{\sigma_{\bar{M}_2} \dagger} \xi_{s_{\bar{M}_2}}^{\sigma^{\prime}_{\bar{M}_2}} = \delta_{\sigma_{\bar{M}_{2}} \sigma^{\prime}_{\bar{M}_{2}}}$, we observe that
\begin{equation}
\sum_{m_{l}, m_{s}} C_{s_{M_{1}}, s_{\bar{M}_{2}}, s}^{\sigma_{M_{1}}, \sigma_{\bar{M}_{2}}, m_{s}} C_{l, s, J}^{m_{l}, m_{s}, m_{J}} Y_{l}^{m_{l}}(\bm{\hat{x}}) = \xi_{s_{M_1}}^{\sigma_{M_1} \dagger} \xi_{s_{\bar{M}_2}}^{\sigma_{\bar{M}_2} \dagger} \mathcal{Y}_{l, s}^{J, m_{J}}(\bm{\hat{x}}),
\end{equation}
which can be inserted in \eqref{plaexp} to obtain
\begin{multline}
\delta_{j j^{\prime}} \delta_{\sigma_{M_{1}^{(j)}} \sigma_{M_{1}^{(j^{\prime})}}} \delta_{\sigma_{\bar{M}_{2}^{(j)}} \sigma_{\bar{M}_{2}^{(j^{\prime})}}} e^{i \bm{p}^{(j^{\prime})} \cdot \bm{r}} = \\
\xi_{s_{M_{1}^{(j)}}}^{\sigma_{M_{1}^{(j)}}\dagger} \xi_{s_{\bar{M}_{2}^{(j)}}}^{\sigma_{\bar{M}_{2}^{(j)}}\dagger}
\Biggl( 4 \pi \sum_{J, m_{J}} \sum_{l^{(j)}, s^{(j)}} \sum_{l^{(j^{\prime})}, s^{(j^{\prime})}}
\delta_{j j^{\prime}} \delta_{l^{(j)} l^{(j^{\prime})}} \delta_{s^{(j)} s^{(j^{\prime})}} i^{l^{(j)}} j_{l^{(j)}}(p^{(j)} r) \mathcal{Y}_{l^{(j)}, s^{(j)}}^{J, m_{J}}(\bm{\hat{r}}) \mathcal{Y}_{l^{(j^{\prime})}, s^{(j^{\prime})}}^{J, m_{J} \dagger}(\bm{\hat{p}}) \Biggr)
\xi_{s_{M_{1}^{(j^{\prime})}}}^{\sigma_{M_{1}^{(j^{\prime})}}} \xi_{s_{\bar{M}_{2}^{(j^{\prime})}}}^{\sigma_{\bar{M}_{2}^{(j^{\prime})}}}
\label{vecplaexp}
\end{multline}
and in \eqref{ampexp} to obtain
\begin{multline}
f^{j \leftarrow j^{\prime}}_{\sigma_{M_{1}^{(j)}}, \sigma_{\bar{M}_{2}^{(j)}}; \sigma_{M_{1}^{(j^{\prime})}}, \sigma_{\bar{M}_{2}^{(j^{\prime})}}}(\bm{\hat{p}} \cdot \bm{\hat{r}}) = \\
\xi_{s_{M_{1}^{(j)}}}^{\sigma_{M_{1}^{(j)}}\dagger} \xi_{s_{\bar{M}_{2}^{(j)}}}^{\sigma_{\bar{M}_{2}^{(j)}}\dagger}
\Biggl( 4 \pi \sum_{J, m_{J}} \sum_{l^{(j)}, s^{(j)}} \sum_{l^{(j^{\prime})}, s^{(j^{\prime})}}
f_{J; l^{(j)},s^{(j)}; l^{(j^{\prime})},s^{(j^{\prime})}}^{j \leftarrow j^{\prime}} \mathcal{Y}_{l^{(j)}, s^{(j)}}^{J, m_{J}}(\bm{\hat{r}}) \mathcal{Y}_{l^{(j^{\prime})}, s^{(j^{\prime})}}^{J, m_{J} \dagger}(\bm{\hat{p}}) \Biggr)
\xi_{s_{M_{1}^{(j^{\prime})}}}^{\sigma_{M_{1}^{(j^{\prime})}}} \xi_{s_{\bar{M}_{2}^{(j^{\prime})}}}^{\sigma_{\bar{M}_{2}^{(j^{\prime})}}}.
\label{vecampexp}
\end{multline}
Finally, inserting Eqs.~\eqref{vecplaexp}, \eqref{vecampexp} in \eqref{scadef} yields the desired expansion:
\begin{equation}
\psi_{\bm{\hat{p}}; \sigma_{M_{1}^{(j)}}, \sigma_{\bar{M}_{2}^{(j)}}; \sigma_{M_{1}^{(j^{\prime})}}, \sigma_{\bar{M}_{2}^{(j^{\prime})}}}^{j \leftarrow j^{\prime}}(\bm{r}) = \xi_{s_{M_{1}^{(j)}}}^{\sigma_{M_{1}^{(j)}}\dagger} \xi_{s_{\bar{M}_{2}^{(j)}}}^{\sigma_{\bar{M}_{2}^{(j)}}\dagger}
\Biggl(\sum_{J, m_{J}} \sum_{l^{(j^{\prime})}, s^{(j^{\prime})}} \psi_{J, m_{J}; l^{(j^{\prime})},s^{(j^{\prime})}}^{j \leftarrow j^{\prime}}(\bm{r}) \mathcal{Y}_{l^{(j^{\prime})}, s^{(j^{\prime})}}^{J, m_{J} \dagger}(\bm{\hat{p}}) \Biggr) \xi_{s_{M_{1}^{(j^{\prime})}}}^{\sigma_{M_{1}^{(j^{\prime})}}} \xi_{s_{\bar{M}_{2}^{(j^{\prime})}}}^{\sigma_{\bar{M}_{2}^{(j^{\prime})}}}
\end{equation}
with
\begin{multline}
\psi_{J, m_{J}; l^{(j^{\prime})},s^{(j^{\prime})}}^{j \leftarrow j^{\prime}}(\bm{r}) \simeq \\
\frac{1}{r} \sqrt{\frac{2}{\pi} \frac{\mu^{(j)}}{p^{(j)}}} \sum_{l^{(j)}, s^{(j)}} i^{l^{(j)}}
\Biggl[\delta_{j j^{\prime}} \delta_{l^{(j)} l^{(j^{\prime})}} \delta_{s^{(j)} s^{(j^{\prime})}} \sin\biggl(p^{(j)}r - l^{(j)}\frac{\pi}{2}\biggr)
+p^{(j)} f_{J; {l^{(j)},s^{(j)}; l^{(j^{\prime})},s^{(j^{\prime})}}}^{j \leftarrow j^{\prime}} e^{i(p^{(j)} r - l^{(j)}\frac{\pi}{2})} \Biggr] \mathcal{Y}_{l^{(j)}, s^{(j)}}^{J, m_{J}}(\bm{\hat{r}})
\label{partwav}
\end{multline}
(compare for example with Eqs.~(15.12), (15.16) in \cite{New82}).

The total unpolarized cross section
\begin{equation}
\sigma^{j \leftarrow j^{\prime}} = \frac{1}{2 s_{M_{1}^{(j^{\prime})}} + 1} \frac{1}{2 s_{\bar{M}_{2}^{(j^{\prime})}} + 1} \sum_{\sigma_{M_{1}^{(j)}}, \sigma_{\bar{M}_{2}^{(j)}}} \sum_{\sigma_{M_{1}^{(j^{\prime})}}, \sigma_{\bar{M}_{2}^{(j^{\prime})}}}
\int \mathrm{d}\bm{\hat{r}}
\Bigl\lvert f^{j \leftarrow j^{\prime}}_{\sigma_{M_{1}^{(j)}}, \sigma_{\bar{M}_{2}^{(j)}}; \sigma_{M_{1}^{(j^{\prime})}}, \sigma_{\bar{M}_{2}^{(j^{\prime})}}}(\bm{\hat{p}} \cdot \bm{\hat{r}}) \Bigr\rvert^{2}
\label{csecgen}
\end{equation}
(see for example Eq.~(5.9) in \cite{Tay72}) can be calculated from the scattering states \eqref{partwav}, this is, in terms of the corresponding scattering amplitudes. Concretely, inserting \eqref{ampexp} into \eqref{csecgen}, after a lengthy calculation one eventually finds
\begin{equation}
\sigma^{j \leftarrow j^{\prime}} = \frac{1}{2 s_{M_{1}^{(j^{\prime})}} + 1} \frac{1}{2 s_{\bar{M}_{2}^{(j^{\prime})}} + 1} \sum_{J} 4 \pi (2 J + 1) \sum_{l^{(j)}, s^{(j)}} \sum_{l^{(j^{\prime})}, s^{(j^{\prime})}} \Bigl\lvert f_{J; l^{(j)},s^{(j)}; l^{(j^{\prime})},s^{(j^{\prime})}}^{j \leftarrow j^{\prime}}\Bigr\rvert^{2}.
\end{equation}
If one further imposes conservation of parity and $C$-parity, then it is ultimately possible to write the total unpolarized cross section as a sum of $J^{PC}$ cross sections, $\sigma^{j \leftarrow j^{\prime}} = \sum_{J^{PC}} \sigma^{j \leftarrow j^{\prime}}_{J^{PC}}$, with each contribution involving only $(l, s)$ partial waves coupling to the specific $J^{PC}$ quantum numbers,
\begin{equation}
\sigma^{j \leftarrow j^{\prime}}_{J^{PC}} = \frac{4 \pi (2 J + 1)}{(2 s_{M_{1}^{(j^{\prime})}} + 1) (2 s_{\bar{M}_{2}^{(j^{\prime})}} + 1)} \sum_{k} \sum_{k^{\prime}} \Bigl\lvert f_{J; l^{(j)}_{k},s^{(j)}_{k}; l^{(j^{\prime})}_{k^{\prime}},s^{(j^{\prime})}_{k^{\prime}}}^{j \leftarrow j^{\prime}} \Bigr\rvert^{2}
\end{equation}
where $k^{\prime}$ and $k$ label distinct $(l, s)$ partial waves coupling to $J^{PC}$ in the initial and final channel respectively.

\twocolumngrid

\bibliography{scatbib}

\end{document}